\documentclass{ws-rv9x6}

\usepackage{subfigure}   
\usepackage{ws-rv-thm}   
\usepackage{ws-rv-van}   
\usepackage{booktabs}	 
\usepackage{bm}
\usepackage{color}
\usepackage[colorlinks]{hyperref}
\makeindex

\definecolor{darkblue}{rgb}{0,0,0.3}
\hypersetup{citecolor=darkblue}
\hypersetup{linkcolor=darkblue}

\newcommand{\be}{\begin{equation}}
\newcommand{\ee}{\end{equation}}
\renewcommand{\c}{\,,}
\newcommand{\p}{\,.}
\def\varM{\mathcal{M}}
\def\calO{\mathcal{O}}
\def\ud{\mathrm{d}}
\def\ui{\mathrm{i}}
\def\n{\nabla\!}

\begin{document}

\chapter{Theory of Gravitational Waves}

\author[Alexandre Le Tiec, J\'er\^ome Novak]{\vspace{-0.1cm}Alexandre Le Tiec and J\'er\^ome Novak\footnote{To be published in the book \textit{An overview of gravitational waves: Theory and detection}, eds. G. Auger and E. Plagnol (World Scientific, 2016).}}

\address{LUTH, Observatoire de Paris, PSL Research University, \\
		 CNRS, Universit\'e Paris Diderot, Sorbonne Paris Cit\'e, \\
		 \vspace{-0.03cm} 5 place Jules Janssen, 92195 Meudon Cedex, France}

\begin{abstract}
The existence of gravitational radiation is a natural prediction of any relativistic description of the gravitational interaction. In this chapter, we focus on gravitational waves, as predicted by Einstein's general theory of relativity. First, we introduce those mathematical concepts that are necessary to properly formulate the physical theory, such as the notions of manifold, vector, tensor, metric, connection and curvature. Second, we motivate, formulate and then discuss Einstein's equation, which relates the geometry of spacetime to its matter content. Gravitational waves are later introduced as solutions of the linearized Einstein equation around flat spacetime. These waves are shown to propagate at the speed of light and to possess two polarization states. Gravitational waves can interact with matter, allowing for their direct detection by means of laser interferometers. Finally, Einstein's quadrupole formulas are derived and used to show that nonspherical compact objects moving at relativistic speeds are powerful gravitational wave sources.
\end{abstract}

\body

\section{Introduction}\label{s:th_intro}

Together with black holes and the expansion of the Universe, the existence of gravitational radiation is one of the key predictions of Einstein's general theory of relativity.\cite{Ei.16,Ei.18} The discovery of the binary pulsar PSR B1913+16,\cite{HuTa.75} and the subsequent observation of its orbital decay, as well as that of other binary pulsars, have provided strong evidence for the existence of gravitational waves.\cite{WeHu.16,Lo.08} These observations have triggered an ongoing international effort to detect gravitational waves directly, mainly by using kilometer-scale laser interferometric antennas such as the LIGO and Virgo detectors.\cite{Aa.al.15,Ac.al.15}

During the months of September and October 2015, the Advanced LIGO antennas have detected, for the first time, gravitational waves generated by two distinct cosmic sources. These waves were emitted, more than a billion years ago, during the coalescence of two binary black hole systems of $65M_\odot$ and $22M_\odot$, respectively.\cite{Ab.al2.16,Ab.al3.16} Many more gravitational-wave observations are expected to follow before the end of this decade.\cite{Ab.al.16} These are truly exciting times, because the direct observation of gravitational waves is going to have a tremendous impact on physics, astrophysics and cosmology.\cite{SaSc.09}

In this chapter, we provide a short but self-contained introduction to the theory of gravitational waves. No prior knowledge of general relativity shall be assumed, and only those concepts that are necessary for an introductory discussion of gravitational radiation will be introduced. For more extensive treatments, the reader is referred to the resource letter \refcite{Ce.03}, the review articles \refcite{Th.87,ScRi.01,FlHu.05,Bu.07,Bl.14,BuSa.15}, and the topical books \refcite{Mag,CrAn}. Most general relativity textbooks include a discussion of gravitational radiation,
such as Refs.~\refcite{Wei,MTW,Wal,Har,Car,Schu,Str}.

The remainder of this chapter is organized as follows. Section \ref{s:intro_gw} provides a qualitative introduction to gravitational waves. Section \ref{s:geo} introduces the geometrical setting (manifold, metric, connection) that is required to formulate the general theory of relativity, the topic of Sec.~\ref{s:Einstein}. Then, gravitational waves are defined, in Sec.~\ref{s:def_gw}, as solutions of the linearized Einstein equation around flat (Minkowski) spacetime. These waves are shown to propagate at the speed of light and to possess two polarization states. The interaction of gravitational waves with matter, an important topic that underlies their direct detection, is addressed in Sec.~\ref{s:th_inter}. Finally, Sec.~\ref{s:gen} provides an overview of the generation of gravitational radiation by matter sources. In particular, Einstein's quadrupole formulas are used to show, using order-of-magnitude estimates, that nonspherical compact objects moving at relativistic speeds are powerful gravitational wave emitters.

Throughout this chapter we use units in which $c = 1$, except in Secs.~\ref{s:intro_gw} and \ref{s:gen}, where we keep all occurences of the speed of light. Our conventions are those of Ref.~\refcite{MTW}; in particular, we use a metric signature $-,+,+,+$.

\section{What is a Gravitational Wave?}\label{s:intro_gw}

We start with a qualitative discussion of gravitational waves. The existence of gravitational radiation is first shown to be a natural consequence of any relativistic description of the gravitational interaction. Then, the properties of gravitational waves, as predicted by the general theory of relativity, are contrasted with those of electromagnetic waves.

\subsection{Newtonian gravity}\label{ss:Newton}

Among the four known fundamental interactions in Nature, gravitation
was the first to be discovered, described and modeled. Isaac Newton's law
of universal gravitation, first published in 1687, states that two
pointlike massive bodies attract each other through a force $\vec{F}$ whose
norm $\Vert \vec{F} \Vert = G m_1 m_2 / r^2$ is proportional to their masses $m_1$ and
$m_2$, and inversely proportional to the square of
their separation $r$, with $G$ a universal constant. Recalling that this
force derives from a local potential $\Phi$, a common
form of Newton's law is Poisson's equation
\be\label{e:Poisson}
  \nabla^2 \Phi = 4 \pi G \rho \c
\ee
with $\rho$ the mass density of matter, acting as the source of
the gravitational potential $\Phi$. Hence, in Newtonian gravity, the gravitational
interaction acts \emph{instantaneously}. This was
already of some concern to Newton himself, but it clearly became
a significant problem with the advent of Einstein's theory of special
relativity.

\subsection{Special relativity}\label{ss:SR}

In 1887, Abraham Michelson and Edward Morley performed an experiment that was designed to detect the relative motion of matter with respect to the luminiferous \ae{}ther, the hypothetical medium that James Clerk Maxwell introduced to explain the propagation of electromagnetic waves. By making use of what is now called a Michelson interferometer (see Chaps.~3 and 4), Michelson and Morley measured the velocity of light from a common source along two orthogonal directions.

The result of this experiment was negative, as it yielded the same value for the speed of light, irrespective of the position and motion of the Earth around the Sun. This opened up a major problem in physics, whose resolution triggered the formulation, in 1905, of the (special) theory of relativity. Einstein's theory builds upon the following two postulates:
\begin{enumerate}
	\item \emph{Principle of relativity}: the equations describing the laws of physics have the same form in all inertial reference frames; \vspace{0.1cm}
	\item \emph{Invariant light speed}: in a vacuum, light propagates at a constant speed $c$, irrespective of the state of motion of the source.
\end{enumerate}
While the principle of relativity was already realized in Galilean and Newtonian mechanics, the second postulate was responsible for a drastic revision in our understanding of space and time themselves.

One central concept that underlies special relativity is that of spacetime interval between two events. Let $\Delta t$, $\Delta x$, $\Delta y$ and $\Delta z$ denote the coordinate differences between
two events $p$ and $q$ with respect to a global inertial frame of
reference. Then, the \emph{spacetime interval} between those events is
\be\label{e:interval}
 	\Delta s^2 \equiv  - c^2 (\Delta t)^2 + (\Delta x)^2 + (\Delta y)^2 + (\Delta z)^2 \p 
\ee
The form of the interval \eqref{e:interval} is quadratic in the differences of the coordinates, and invariant under the Poincar\'e group\cite{Gou2} (translations, rotations, boosts), thus ensuring that the speed of light is indeed the same in all inertial frames. This observation suggests that, in full analogy with the Euclidean geometry of three-dimensional space, special relativity can be formulated as a theory of the \emph{Lorentzian geometry} of four-dimensional spacetime.

\begin{figure}
	\begin{center}
		\includegraphics[width=13cm]{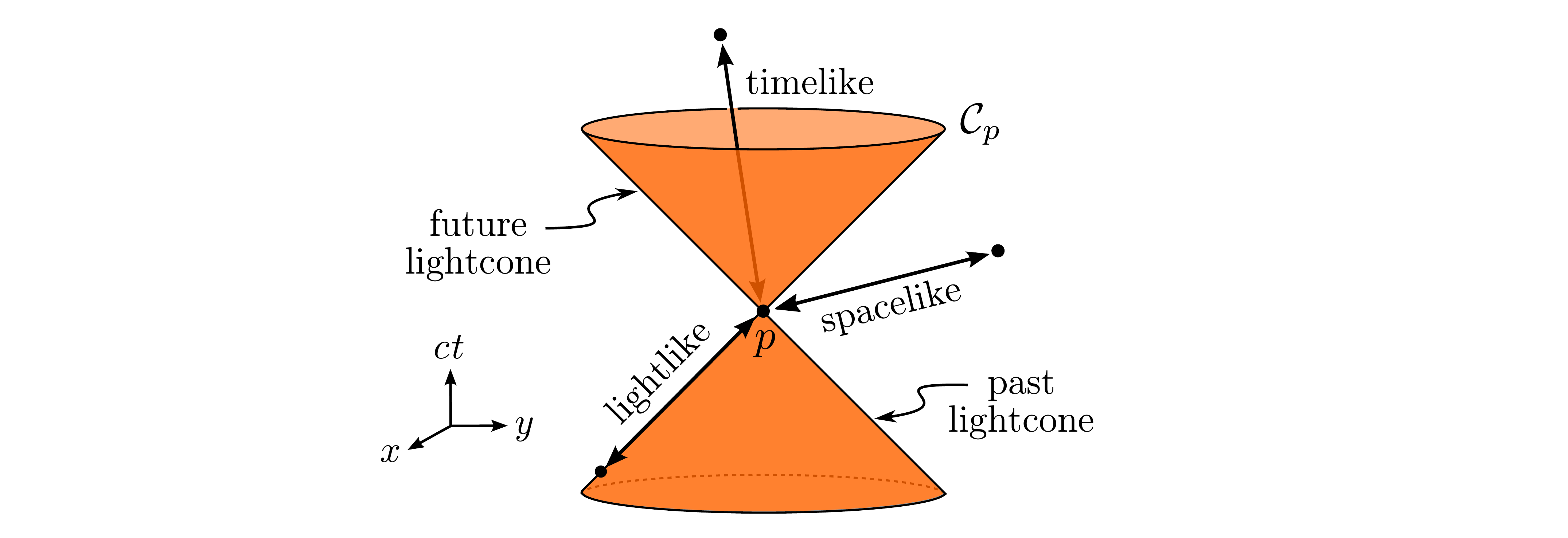}
		\caption{In special relativity, the causal structure of spacetime defines a notion of lightcone $\mathcal{C}_p$ at any event $p$. All events on $\mathcal{C}_p$ are lightlike related to $p$, while all events within (respectively, outside) $\mathcal{C}_p$ are timelike (respectively, spacelike) related to $p$.}
		\label{f:lightcone}
	\end{center}
\end{figure}

Moreover, the spacetime interval can be used to explore the causal structure of spacetime; see \fref{f:lightcone}. Given an event $p$, the \emph{lightcone} $\mathcal{C}_p$ is the set of all events $q$ such that $\Delta s^2 = 0$. These events are said to be \emph{lightlike} related to $p$ because all of them can be reached by a light ray going through $p$. All the events within $\mathcal{C}_p$ are such that $\Delta s^2 < 0$. Those events are said to be \textit{timelike} related to $p$ because a massive particle going through $p$ can, at least in principle, reach any one of them. The remaining events, i.e., the events outside $\mathcal{C}_p$, are such that $\Delta s^2 > 0$. Those events are said to be \textit{spacelike} related to $p$ because no massive particle, nor any light ray going through $p$, can ever reach them. Two events that are spacelike related cannot have any causal influence over each other.

\vspace{0.2cm}

\subsection{Relativistic gravity?}\label{ss:SR_gravity}

Special relativity is the relevant framework to describe the
electromagnetic, weak and strong interactions. Therefore, a natural question is whether the gravitational interaction can be accomodated to ``fit that mold'' as well? A straightforward relativistic extension of Poisson's equation
\eqref{e:Poisson} is to replace the elliptic Laplace operator $\nabla^2$ by the hyperbolic d'Alembert operator and the mass density $\rho$ by a Lorentz covariant source. Hence, one is naturally led to postulate a gravitational field
equation of the form
\be\label{e:scalar_SR}
  \Box \Phi = - \frac{4 \pi G}{c^2} \, T \c
\ee
where $\Box \equiv - \frac{1}{c^2} \frac{\partial^2}{\partial t^2} + \nabla^2$ is the usual
flat-space wave operator, and $T$ is the trace of the energy-momentum
tensor of matter (see \sref{ss:Tmunu}). Such a scalar theory of
gravity obeys the principle of special relativity, and it reproduces
Poisson's equation \eqref{e:Poisson} in the nonrelativistic limit where $c^{-1} \to
0$. However, it disagrees with observations, as it predicts no deflection of
light and the wrong perihelion advance for Mercury.\cite{Ra.04}

Nevertheless, this failed attempt illustrates one central idea behind
any relativistic theory of gravity, namely the requirement to
incorporate a finite velocity for the propagation of the gravitational
interaction. Then, just like in electromagnetism, the propagation of
gravitation at a finite speed should manifest itself through traveling
waves. The notion of a \emph{gravitational wave} thus appears to be a
natural byproduct of any relativistic theory of gravity.

\subsection{Gravitational waves vs electromagnetic waves}

Although gravitational waves and electromagnetic waves share some
similarities, they also differ strongly in their very nature and main
caracteristics: while electromagnetic waves are nothing but
oscillations in the electromagnetic field that propagate \textit{in}
spacetime, gravitational waves ---as predicted by Einstein's general
relativity--- are tiny propagating ripples in the curvature
\textit{of} spacetime itself.

Electromagnetic radiation is produced by the motion of a large number
of microscopic charges, giving rise to an incoherent superposition of
waves with a dipolar structure in the wave zone. Because the wavelengths of
electromagnetic waves are typically much smaller than the size of their
sources, these waves can be used to produce images. Gravitational
radiation, on the other hand, is produced by the bulk motion of
macroscopic masses, giving rise to a coherent superposition of waves
with a quadrupolar structure in the wave zone. Since the wavelengths of gravitational
waves are typically larger than the size of their sources, these waves cannot
be used to produce images; rather their two polarization states are more
akin to ``stereo sound'' information.

Electromagnetic waves interact strongly with matter, and are typically
scattered many times as they propagate away from the sources. This
strong interaction ensures that the power in the field, which decays
like the inverse distance squared to the source, can easily be
detected. Gravitational waves, on the contrary, barely interact
with matter and propagate almost freely in the Universe, thus making
their detection quite chalenging. However, their typical frequency is
low enough that the amplitude of the wave itself, which decays like
the inverse distance, can be tracked in time.

These multiple differences, summarized in Table \ref{tab:comparison}, imply that electromagnetic waves and gravitational waves are complementary sources of information about their astrophysical sources. The forthcoming \emph{multi-messenger astronomy} will soon built upon that complementarity.\cite{An.al.13} In particular, the observation of electromagnetic counterparts to gravitational waves signals could improve our understanding of the progenitors of gamma-ray bursts\cite{Am.al3.13} and core-collapse supernov\ae{}.\cite{Ot.09} More generally, forthcoming gravitational-wave detections will provide the opportunity for multi-messenger analyses, combining gravitational wave with electromagnetic, cosmic ray or neutrino observations.

\begin{table}[h!]
	\tbl{Comparison of the main characteristics of electromagnetic waves and gravitational waves. \vspace{0.1cm}}
	{\begin{tabular}{@{}lll@{}}
		\toprule
					& Electromagnetic waves & Gravitational waves \\
		\midrule
		Nature		& electromagnetic field	& spacetime curvature \\ \noalign{\medskip}
		Sources		& accelerated charges	& accelerated masses \\ \noalign{\medskip}
		Wavelength	& $\ll$ size of source	& $\gtrsim$ size of source \\ \noalign{\medskip}
		Structure	& dipolar				& quadrupolar \\ \noalign{\medskip}
		Coherence	& low					& high \\ \noalign{\medskip}
		Interaction & strong				& weak \\ \noalign{\medskip}
		Detection	& power					& amplitude \\ \noalign{\medskip}
		Analogy		& vision				& audition \\
		\bottomrule
	\end{tabular}}
	\label{tab:comparison}
\end{table}

\section{Gravity is Geometry}\label{s:geo}

In order to properly understand the nature of gravitational waves, we
must first introduce the general theory of relativity. We start from
its geometrical setting, which is motivated by the observation of the
universality of free fall.

\subsection{Equivalence principle}

One key feature singles out gravity from the other fundamental interactions: the observation of the \emph{universality} of free fall. Indeed, all bodies are affected by gravity and, in fact, all bodies fall with the exact same acceleration in an external gravitational field. Thus, the motion of a freely falling body is independent of its mass, and even more remarkably, of its composition. This fact has no natural explanation in Newtonian gravity, where it is assumed that, for all bodies, the inertial mass is, for some mysterious reason, exactly equal to the gravitational mass (or gravitational charge).

Put differently, a gravitational field is, locally, equivalent to an accelerated reference frame, as the following \textit{Gedankenexperiment} illustrates: a freely falling observer in a freely falling lift cannot determine, by any local experiment, the possible existence of an external gravitational field. While devising his relativistic theory of gravitation, Einstein relied crucially upon this so-called ``equivalence principle.'' Nowadays, physicists distinguish the three following equivalence principles:
\begin{itemize}
	\item \emph{Weak equivalence principle}: given the same initial
  position and velocity, all test bodies fall along
  the same trajectories. \vspace{0.15cm}
	\item \emph{Einstein equivalence principle}: in a local inertial
  frame, all nongravitational laws of physics are given by their
  special-relativistic form. \vspace{0.15cm}
	\item \emph{Strong equivalence principle}: it is always possible to
  remove the effects of an exterior gravitational field by choosing
  a local inertial frame in which all the laws of physics, including those of gravity,
  take the same form as in the absence of this exterior
  gravitational field.
\end{itemize}
Whereas the strong equivalence principle implies the Einstein equivalence principle, which itself implies the weak equivalence principle, none of the converse implications is necessarily true. However, Schiff's conjecture states than any ``reasonable'' theory of gravity which obeys the weak equivalence principle must also obey the Einstein equivalence principle. While all metric theories of gravity obey the weak equivalence principle, general relativity is one of the few such theories that obeys the strong equivalence principle.\cite{Wi.14}

The weak equivalence principle has been tested by various experiments, starting with the historic torsion-balance studies by E\H{o}tv\H{o}s,\cite{Eo.al.22} which already reached a relative accuracy of $10^{-8}$. Several recent experiments used similar setups, but achieved the remarkable upper limit of $10^{-13}$ on the violation of the weak equivalence principle.\cite{Sc.al2.08} The MICROSCOPE mission will test the weak equivalence principle in space\cite{To.al.12} down to an accuracy of $10^{-15}$. The strong equivalence principle has been tested using lunar laser ranging\cite{Mu.al.12} and binary pulsar timing,\cite{Go.al.11} with constraints on the Nordtvedt parameter $\eta$ and the $\Delta$ parameter at the $3.6 \times 10^{-4}$ and $4.6 \times 10^{-3}$ levels, respectively. Future studies of a recently discovered pulsar\cite{Ra.al.14} in a triple system with two white dwarfs will soon provide new tests of the strong equivalence principle.

\begin{figure}
	\begin{center}
		\includegraphics[width=7.5cm]{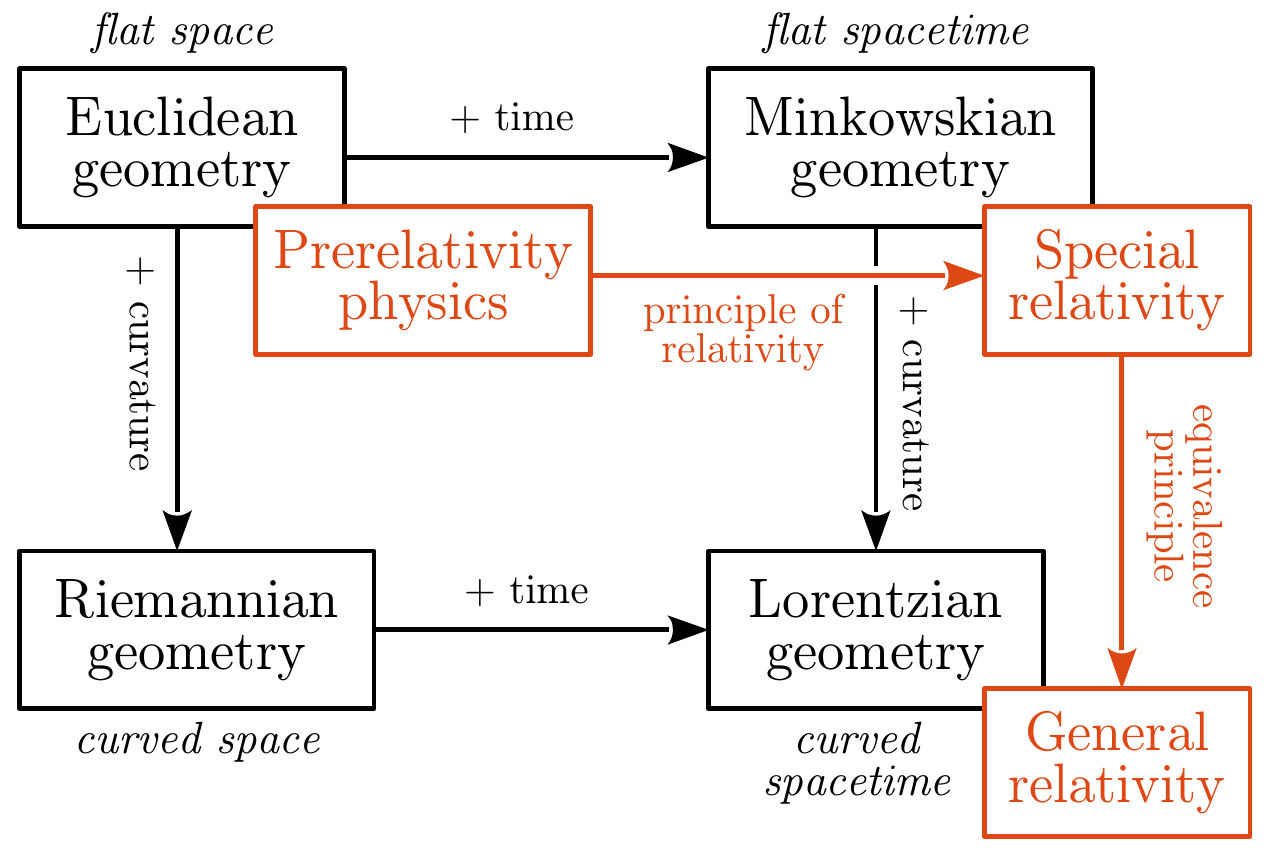}
		\caption{The historical genesis of the general theory of relativity required a combination of physical principles and geometrical concepts.}
		\label{f:geo}
	\end{center}
\end{figure}

According to the equivalence principle, the paths of freely falling bodies define a preferred set of curves in spacetime. \!\!This suggests that gravitation is not a property of matter but, rather, a feature of the structure of spacetime itself. \!Indeed, following a decade-long struggle Einstein realized that gravity can in fact be understood as the manifestation of the curvature of spacetime. This conceptual breakthrough requires ``only'' a simple generalization of the flat Lorentzian geometry of special relativity (i.e. of Minkowski's spacetime) to a curved Lorentzian geometry, just like the flat Euclidean geometry can be generalized to a curved Riemannian geometry; see Fig.~\ref{f:geo}. Thus, general relativity is a theory of the structure of space and time that accounts for all the physical effects of gravity in terms of the curvature of the geometry of spacetime. It turns out that the mathematical concept be suited to describe such a smooth set of points is that of manifold.

\subsection{Notion of manifold}

\begin{figure}
	\begin{center}
		\includegraphics[width=7.5cm]{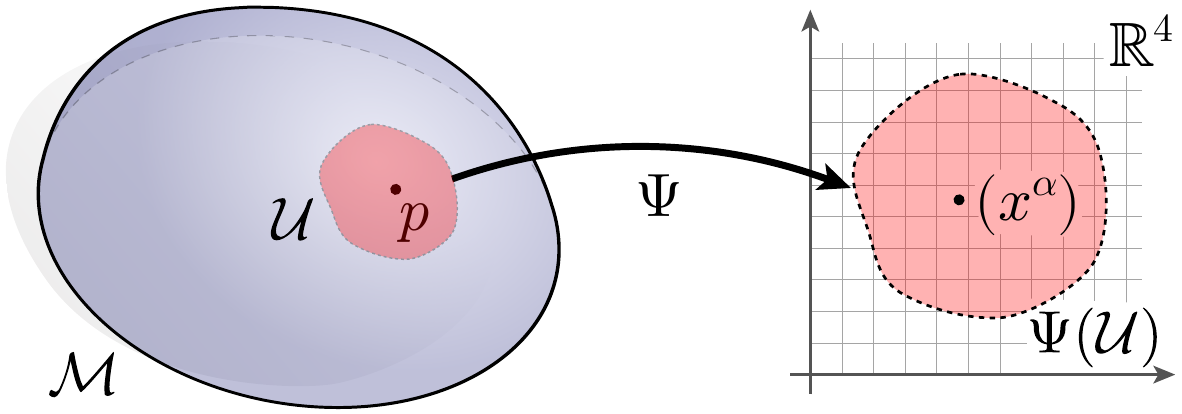}
		\caption{Over a four-dimensional manifold $\mathcal{M}$, a neighbourhood $\mathcal{U}$ of a point $p \in \mathcal{M}$ can be mapped to a subset $\Psi(\mathcal{U})$ of $\mathbb{R}^4$. (Only two dimensions are shown.)}
		\label{f:manifold}
	\end{center}
\end{figure}

To begin our exploration of the structure of spacetime, we need the notion of manifold, that is of a set of points (spacetime events) that ``looks locally'' like the set $\mathbb{R}^4$ of quadruplets of real numbers, but whose global properties may differ from those of $\mathbb{R}^4$. More precisely, a four-dimensional \emph{manifold} $\varM$ is a topological space such that, at every point, it is possible to define a local neighbourhood that is isomorphic to an open set of $\mathbb{R}^4$. Loosely speaking, this means that for a ``sufficiently small'' part of $\varM$, it is possible to assign four numbers, called coordinates, to every point $p$. Therefore, a \emph{coordinate system} (or \emph{chart}) over an open subset $\mathcal{U}$ of $\varM$ is a map (see~\fref{f:manifold})
\begin{align}\label{e:def_manifold}
	\Psi: \mathcal{U} \subset \varM & \longrightarrow \Psi(\mathcal{U}) \subset \mathbb{R}^4 \nonumber \\
		  			p \hspace{0.5cm} & \longmapsto (x^0,x^1,x^2,x^3) \p
\end{align}
Hereafter it will prove convenient to use the shorthand $(x^\alpha) \equiv (x^0\!,x^1\!,x^2\!,x^3)$ to denote a coordinate system. It is of uttermost importance to realize that coordinates are by no means unique. The choice of a coordinate system over (part of) a manifold is entirely free, and coordinates are devoid of physical significance.

Some familiar examples of two-dimensional manifolds include the plane, the cylinder, the sphere and the torus. Note that the definition of a manifold is \emph{intrinsic}, in the sense that a manifold needs not be embedded into a higher dimensional space. For instance, the sphere $\mathbb{S}^2$ can be defined without any reference to the Euclidean space $\mathbb{R}^3$.

In general, several charts are needed in order to cover a given manifold. A finite collection of charts $\left( \mathcal{U}_k, \Psi_k\right)$, where $\bigcup_k \mathcal{U}_k = \varM$ is called an \emph{atlas}. A manifold $\varM$ is said to be \emph{differentiable} (or smooth) if, for every non-empty intersection $\mathcal{U}_i \cap \mathcal{U}_j$, the function $\Psi_i \circ \Psi_j^{-1} : \Psi_j(\mathcal{U}_i \cap \mathcal{U}_j) \longrightarrow \Psi_i(\mathcal{U}_i \cap \mathcal{U}_j)$ is differentiable (or smooth).

\vspace{0.3cm}

\subsection{Vectors, dual vectors and tensors}

To formulate the laws of physics in curved spacetime, the notions of scalar field, vector field, etc, need to be generalized to the case of a manifold. The central idea here is the possibility to change the chart, or coordinate system, over the spacetime manifold. Since the laws of physics should not depend on a particular choice of coordinates, their form should be \emph{covariant} under general coordinate transformations. This requirement generalizes the first postulate of special relativity, recalled in \sref{ss:SR}, to all frames of reference; hence the name ``general relativity.'' Physical laws should thus be expressed in terms of mathematical objects that transform in a well-defined manner under general coordinate transformations, i.e., in terms of tensors.

The simplest type of tensor is the \emph{scalar field}, an application $S : \mathcal{M} \to \mathbb{R}$ that associates a real number $S(p)$ to any point $p \in \mathcal{M}$. Under a coordinate transformation $x^\alpha \to x'^\alpha(x)$, a scalar field transforms as
\be
  \label{e:def_scal}
  S'(x') = S(x) \, ,
\ee
where, following a widespread abuse of notation, we implicitly identify each point with its coordinates in a given chart. Well known examples of scalar fields include, for instance, the density and temperature of a fluid.

\subsubsection*{Curves and vectors}

In affine spaces, such as the ordinary three-dimensional space of Euclidean geometry and the four-dimensional spacetime of Minkowskian geometry, a vector is equivalent to a point (once a choice of origin has been made). In the more general case of a manifold, however, this is not true. Nevertheless, a well-defined notion on a manifold is that of curve. Vectors can then be defined as tangent vectors to a given curve.

\begin{figure}
	\begin{center}
		\includegraphics[width=11.5cm]{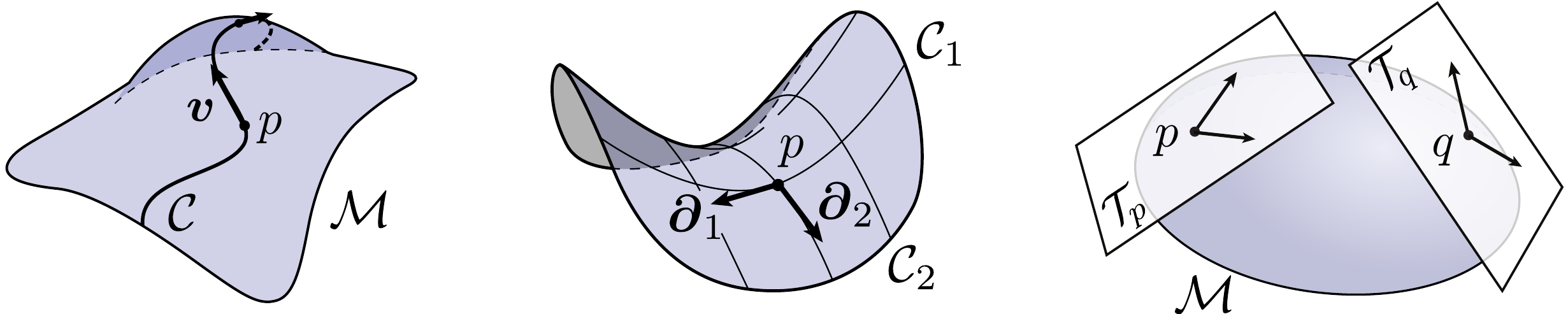}
		\caption{\textsl{Left panel:} a vector $\boldsymbol{v}$ defined as a directional derivative operator at a point $p$ along a curve $\mathcal{C}$. \textsl{Middle panel:} the basis vectors $\boldsymbol{\partial}_\alpha$ tangent to the coordinate lines $\mathcal{C}_\alpha$ associated to a coordinate system $(x^\alpha)$. \textsl{Right panel:} the tangent spaces $\mathcal{T}_p$ and $\mathcal{T}_q$ at two distinct points $p$ and $q$ over a manifold $\mathcal{M}$. (Only two dimensions are shown.)}
		\label{f:vectors}
	\end{center}
\end{figure}

Given a coordinate system $(x^\alpha)$, a \emph{curve} $\mathcal{C}$ is given in parametric form by four equations of the form $x^\alpha = X^\alpha(\lambda)$, with $\lambda \in \mathbb{R}$ the parameter along that curve. Then, the \emph{tangent vector} $\boldsymbol{v}$ to the curve $\mathcal{C}$ at a point $p \in \mathcal{C}$ is the operator that associates to every scalar field $f: \varM \to \mathbb{R}$ its directional derivative along $\mathcal{C}$ (see Fig.~\ref{f:vectors}, left panel):
\be\label{e:def_vector2}
	\boldsymbol{v}(f) \equiv \frac{\ud f}{\ud \lambda}\bigg\vert_\mathcal{C} = \sum_{\alpha=0}^3 \frac{\partial f}{\partial x^\alpha} \, \frac{\ud X^\alpha}{\ud \lambda} \p  
\ee
At every point $p$, there exist four curves associated to the coordinates $(x^\alpha)$: the \emph{coordinates lines} $\mathcal{C}_\alpha$. For all $0 \leqslant \alpha \leqslant 3$, $\mathcal{C}_\alpha$ is the curve parameterized by $\lambda = x^\alpha$, going through $p$, and such that the coordinates $x^\beta$ are constant for all $\beta \neq \alpha$. The tangent vector to the curve $\mathcal{C}_\alpha$ is denoted $\boldsymbol{\partial}_\alpha$, as shown in the middle panel of Fig.~\ref{f:vectors}. From the definition \eqref{e:def_vector2}, its action on a scalar field $f$ reads
\be\label{e:def_d0}
	\boldsymbol{\partial}_\alpha(f) = \frac{\ud f}{\ud x^\alpha}\bigg\vert_\mathcal{C_\alpha}  = \frac{\partial f}{\partial x^\alpha}.
\ee
The tangent vectors to the coordinate lines act on scalar fields by returning their partial derivatives with respect to the coordinates; hence the notation. Combining Eqs.~\eqref{e:def_vector2} and \eqref{e:def_d0}, we obtain a relation that holds for any scalar field $f$, which implies
\be
  \label{e:vector_base}
  \boldsymbol{v} = \sum_{\alpha=0}^3 v^\alpha \, \boldsymbol{\partial}_\alpha \, ,
\ee
where $v^\alpha = \ud X^\alpha / \ud \lambda$ are the components of the vector $\boldsymbol{v}$ with respect to the coordinate basis vectors $\boldsymbol{\partial}_\alpha$.The space spanned by the four basis vectors $\boldsymbol{\partial}_\alpha$ at the point $p$ is a four-dimensional vector space at that point, the \emph{tangent space} $\mathcal{T}_p$. Beware that, in general, there are as many tangent spaces $\mathcal{T}_p$ as there are points $p$ in a manifold $\mathcal{M}$ (see Fig.~\ref{f:vectors}, right panel).

Recall that the choice of coordinates over part of a manifold is arbitrary. It can easily be shown that under a coordinate transformation $x^\alpha \to x'^\alpha(x)$, the components of a vector $\boldsymbol{v}$ transform as
\be\label{e:def_vector}
  v'^\alpha(x') = \sum_{\beta=0}^3 \frac{\partial x'^\alpha}{\partial x^\beta} \, v^\beta(x) \equiv \frac{\partial x'^\alpha}{\partial x^\beta} \, v^\beta(x) \c
\ee
where, in the second equality, we introduced \emph{Einstein's convention} of summation over repeated indices. From now on we will always use this convention to simplify the notations.

\vspace{0.3cm}

\subsubsection*{Dual vectors and tensors}

A fundamental operation on vectors consists in assigning to them a number, and doing so in a linear manner. A \emph{dual vector} is a map
\be
	\boldsymbol{\omega} : \mathcal{T}_p \longrightarrow \mathbb{R}
\ee
that associates a real number to any vector defined at a point $p \in \mathcal{M}$, and such that $\boldsymbol{\omega}(\lambda \boldsymbol{u} + \boldsymbol{v}) = \lambda \, \boldsymbol{\omega}(\boldsymbol{u}) + \boldsymbol{\omega}(\boldsymbol{v})$ for all $\lambda \in \mathbb{R}$ and for all $\boldsymbol{u},\boldsymbol{v} \in \mathcal{T}_p$. The set of all such dual vectors is a four-dimensional vector space on $\mathcal{T}_p$. It is called the \emph{dual space} to $\mathcal{T}_p$ and is denoted $\mathcal{T}_p^*$. Given a basis of $\mathcal{T}_p^*$, any dual vector $\boldsymbol{\omega}$ can be written as a linear combination of these basis dual vectors, with components $\omega_\alpha$. Under a change of coordinates $x^\alpha \to x'^\alpha(x)$, these components transforms as
\be
  \label{e:def_form}
  \omega'_\alpha(x') = \frac{\partial x^\beta}{\partial x'^\alpha} \, \omega_\beta(x) \, .
\ee

With these definitions in hand, we may now introduce the most general notion of tensor. A \emph{tensor} of type $(r,s)$ at a point $p$ is a multilinear map
\be
	\boldsymbol{T} : \underbrace{\mathcal{T}_p^* \times \dots \times \mathcal{T}_p^*}_{\text{$r$ times}} \times \underbrace{\mathcal{T}_p \times \dots \times \mathcal{T}_p}_{\text{$s$ times}} \longrightarrow \mathbb{R}
\ee
that associates a real number to $r$ dual vectors and $s$ vectors. A multilinear map is a map that is linear with respect to each of its arguments. Just like vectors and dual vectors, a tensor can be expressed as a linear combination of basis tensors, given by ``tensor products'' of basis vectors and dual vectors, with $4^{r+s}$ components $T^{\alpha_1\cdots\alpha_r}_{\phantom{\alpha_1\cdots\alpha_r}\beta_1\cdots\beta_s}$. The integer $r+s$ is called the rank of the tensor $\boldsymbol{T}$. Under a change of coordinates $x^\alpha \to x'^\alpha(x)$, the components of a tensor transform as
\be
  \label{e:def_tensor}
  T'^{\alpha_1\cdots\alpha_r}_{\phantom{\alpha_1\cdots\alpha_r}\beta_1\cdots\beta_s}(x') =
  \frac{\partial x'^{\alpha_1}}{\partial x^{\mu_1}} \cdots \frac{\partial
    x'^{\alpha_r}}{\partial x^{\mu_r}}\, \frac{\partial x^{\nu_1}}{\partial
    x'^{\beta_1}} \cdots \frac{\partial x^{\nu_s}}{\partial x'^{\beta_s}} \,
  T^{\mu_1\cdots\mu_r}_{\phantom{\mu_1\cdots\mu_r}\nu_1\cdots\nu_s} (x) \, .
\ee
Then, vectors and dual vectors are tensors of type $(1,0)$ and $(0,1)$, respectively. By convention, a scalar field is a tensor field of type $(0,0)$.

\subsubsection*{Notation}

There are two notations commonly used to denote tensors: the index-free notation, such as $\boldsymbol{T}$, and the index notation, such as $T^{\alpha_1 \cdots \alpha_r}_{\phantom{\alpha_1 \cdots \alpha_r}\beta_1 \cdots \beta_s}$. Following Ref.~\refcite{Wal}, we will mostly use a third notation, the \emph{abstract index notation}, which combines the respective advantages of the two other notations. The idea is to avoid introducing a basis, but to use a notation that mimics the index notation. A tensor $\boldsymbol{T}$ of type $(r,s)$ will then be written $T^{a_1 \cdots a_r}_{\phantom{a_1 \cdots a_r}b_1 \cdots b_s}$, where the Latin indices $a_i$ and $b_j$ do \emph{not} represent components in a given basis. Rather, those indices provide information about the type of a given tensor and the order in which it ``acts'' on dual vectors and vectors. For instance, $T^{ab}_{\phantom{ab}c}$ denotes a tensor of type $(2,1)$ that acts linearly on two dual vectors and a vector. Hereafter, Latin indices $a,b,c,\dots$ from the beginning of the alphabet will be abstract, whereas Greek indices $\alpha,\beta,\gamma,\dots$ will be used for tensor components with respect to a given basis. We will use indices $i,j,k,\dots$ from the second part of the Latin alphabet to denote purely spatial components of a tensor.

\subsection{Metric tensor}

A key concept in vector spaces is that of scalar product. In special relativity, the scalar product $\boldsymbol{u} \cdot \boldsymbol{v}$ between two four-dimensional vectors $\boldsymbol{u}$ and $\boldsymbol{v}$ reads
\be\label{e:SR_scalar_prod}
	\boldsymbol{u} \cdot \boldsymbol{v} \equiv - u^0 v^0 + u^1 v^1 + u^2 v^2 + u^3 v^3 = \eta_{\alpha\beta} \, u^\alpha v^\beta \c 
\ee
where $\eta_{\alpha\beta} = \text{diag} \, (-1,+1,+1,+1)$ denote the components of the \emph{Minkowski metric} $\eta_{ab}$ with respect to global inertial coordinates $(x^\alpha) = (t,x,y,z)$. In special relativity, any two events $p$ and $q$ can always be related by a vector, say $\bm{s}$, with components $s^\alpha = \Delta x^\alpha$. Therefore, the spacetime interval \eqref{e:interval} between those events is nothing but the scalar product $\bm{s} \cdot \bm{s}$.

In general relativity, however, it is not possible to connect any two points on a manifold $\varM$ by a vector. One has to work locally, in the tangent space $\mathcal{T}_p$ at a given point $p$. Thus, at every point $p \in \varM$, one defines a symmetric rank-two tensor $g_{ab}$ that acts linearly on all couples of vectors of $\mathcal{T}_p$, and which is nondegenerate, i.e., such that if $g_{ab} u^a v^b =0$ for all $v^a$, then $u^a = 0$. The scalar product between two vectors $u^a$ and $v^a$ then reads
\be
	g_{ab} u^a v^b = g_{\alpha\beta} u^\alpha v^\beta \c
\ee
where $g_{\alpha\beta}$, $u^\alpha$ and $v^\beta$ denote the components of the tensors $g_{ab}$, $u^a$ and $v^b$ with respect to a given basis. For any $p \in \mathcal{M}$, one can always construct a basis of $\mathcal{T}_p$ such that $g_{\alpha\beta}(p) = \eta_{\alpha\beta}$. The metric is said to have a Lorentzian signature $-,+,+,+$. Any such tensor field $g_{a b}$ is a \emph{metric} on $\varM$, and the couple $(\varM,g_{ab})$ is called a \emph{spacetime}.

Given a coordinate system $(x^\alpha)$ on $\varM$, let $p$ and $q$ be two nearby events with coordinates $(x_0^\alpha)$ and $(x_0^\alpha + \ud x^\alpha)$. If $g_{\alpha\beta}$ denote the components of the metric $g_{ab}$ with respect to the coordinates $(x^\alpha)$, then the \emph{spacetime interval} between $p$ and $q$ is the number
\be\label{e:ds2_GR}
	\ud s^2 = g_{\alpha\beta} \, \ud x^\alpha \ud x^\beta \p
\ee
This interpretation justifies the name ``metric'' given to the tensor $g_{ab}$. Note that, just like the spacetime interval \eqref{e:interval} in special relativity, the spacetime interval \eqref{e:ds2_GR} is not necessarily positive. To determine the interval between two events that are not infinitesimally close, one must first specify a curve connecting those events, and then integrate the line element $(\pm \ud s^2)^{1/2}$ along that curve. The result will, in general, depend on the curve chosen, but not on the coordinate system.

Since $g_{ab}$ is nondegenerate, one can always define the \emph{inverse metric} $g^{ab}$ such that
\be\label{e:inv_metric}
  g^{ab} g_{bc} = \delta^a_{\phantom{a}c} \c
\ee
where $\delta^a_{\phantom{a}c}$ denotes the identity operator from $\mathcal{T}_p$ to $\mathcal{T}_p$. The metric $g_{ab}$ and the inverse metric $g^{ab}$ can be used to ``lower'' and ``raise'' indices on tensors. For instance, through the definition of the scalar product and \eref{e:inv_metric}, $g_{ab}$ and $g^{ab}$ define one-to-one relations between vectors and dual vectors:
\begin{subequations}\label{e:raise_lower}
	\begin{align}
		v_a &\equiv g_{ab} v^b \c \\
		\omega^a &\equiv g^{ab} \omega_b \p
	\end{align}
\end{subequations}
The reason why the distinction between vectors and dual vectors is never made in prerelativity physics is because the components of the Euclidean metric $f_{ab}$ of three-dimensional space with respect to a Cartesian coordinate system simply read $f_{\alpha\beta} = \text{diag}\,(+1,+1,+1)$, such that $v_i = v^i$.

Given a metric, it is possible to define the type of a vector $v^a$ by making use of its norm (squared) $g_{ab} v^a v^b = v^a v_a$. Indeed, by analogy with the types of intervals defined using the lightcones in special relativity (recall \fref{f:lightcone}), a nonzero vector $v^a$ is said to be \emph{spacelike} if and only if $v^a v_a>0$, \emph{timelike} if and only if $v^a v_a<0$, and \emph{lightlike} (or \emph{null}) if and only if $v^a v_a=0$.

\subsection{Covariant derivative}\label{ss:cov_deriv}

Mathematically, the laws of physics are expressed as differential equations. To formulate those laws in a curved spacetime, one needs the notion of the derivative of a vector field (and more generally of a tensor field). Such a notion requires the comparison of two vectors defined at two nearby points $p$ and $q$, and thus the information required to ``connect'' the tangent spaces $\mathcal{T}_p$ and $\mathcal{T}_q$. However, given a manifold $\varM$, there exists an infinite number of such connections. We will see that the existence of a (Lorentzian) metric $g_{ab}$ on a spacetime manifold singles out a unique connection: the Levi-Civita connection. 

Given a manifold $\mathcal{M}$, a \emph{covariant derivative} (or \emph{connection}) is an application $\nabla$ that assigns to any tensor field $T^{a_1 \dots a_r}_{\phantom{a_1 \dots a_r}b_1 \dots b_s}$ of type $(r,s)$ a tensor field $\n_c T^{a_1 \dots a_r}_{\phantom{a_1 \dots a_r}b_1 \dots b_s}$ of type $(r,s+1)$, and which obeys the usual properties satisfied by a derivative operator: linearity, Leibniz rule, etc. In particular, we require the condition of \emph{absence of torsion}, i.e., that for a scalar field $S$,
\begin{equation}\label{e:no_torsion}
  \n_a \n_b S = \n_b \n_a S \p
\end{equation}

The covariant derivative of a scalar field $S$, denoted $\n_aS$, is of course a field of dual vectors. As expected, its components with respect to a coordinate basis are simply the partial derivatives with respect to the coordinates, such that
\be
	\n_\alpha S = \frac{\partial S}{\partial x^\alpha} \p
\ee
Using the chain rule, this is consistent with the law \eqref{e:def_form} of transformation of the components of a dual vector.

But how does a covariant derivative act on tensors of higher rank? Given a field of basis vectors $(\boldsymbol{e}_\alpha)$, the action of a connection $\nabla$ on a vector field $\bm{v} = v^\beta \bm{e}_\beta$ can easily be written down. In components, one finds
\be\label{e:deriv_vect}
	\n_\alpha \boldsymbol{v} = \n_\alpha \big( v^\beta \boldsymbol{e}_\beta \bigr) = \frac{\partial v^\beta}{\partial x^\alpha} \, \boldsymbol{e}_\beta + v^\beta \n_\alpha \boldsymbol{e}_\beta \c
\ee
where we used the fact that each component $v^\beta$ is a scalar field on $\mathcal{M}$. Then, to fully specify the derivative operator $\nabla$, one must specify a number of scalar fields, the \emph{connection coefficients} $C^\gamma_{\phantom{\gamma}\alpha\beta}$, such that
\be
	\n_\alpha \boldsymbol{e}_\beta = C^\gamma_{\phantom{\gamma}\alpha\beta} \, \boldsymbol{e}_\gamma \p
\ee
Replacing this expression into \eref{e:deriv_vect}, and exchanging the dummy indices $\beta$ and $\gamma$ in the second term, the components $\n_\alpha v^\beta$ of the tensor field $\n_a v^b$ of type $(1,1)$ simply read
\be
	\n_\alpha v^\beta = \partial_\alpha v^\beta + C^\beta_{\phantom{\gamma}\alpha\gamma} \, v^\gamma \c
\ee
where we introduced the notation $\partial_\alpha v^\beta \equiv \partial v^\beta / \partial x^\alpha$ for the ordinary partial derivative. A similar expression for the covariant derivative of a dual vector field can easily be established, and extended to the general case of a type $(r,s)$ tensor field, for which
\begin{align}\label{e:der_tensor}
	\n_\gamma T^{\alpha_1\cdots\alpha_r}_{\phantom{\alpha_1\cdots\alpha_r}\beta_1\cdots\beta_s} &= \partial_\gamma T^{\alpha_1\cdots\alpha_r}_{\phantom{\alpha_1\cdots\alpha_r}\beta_1\cdots\beta_s} + \sum_{i=1}^r C^{\alpha_i}_{\phantom{\alpha_i}\gamma\sigma} \, T^{\alpha_1\cdots\sigma\cdots\alpha_r}_{\phantom{\alpha_1\cdots\sigma\cdots\alpha_r}\beta_1 \cdots \beta_s} \nonumber \\ &\qquad\quad - \sum_{j=1}^s C^\sigma_{\phantom{\sigma}\gamma\beta_j} \, T^{\alpha_1\cdots\alpha_r}_{\phantom{\alpha_1\cdots\alpha_r}\beta_1\cdots\sigma\cdots\beta_s} \p
\end{align}
\begin{figure}[h!]
	\begin{center}
		\includegraphics[width=9cm]{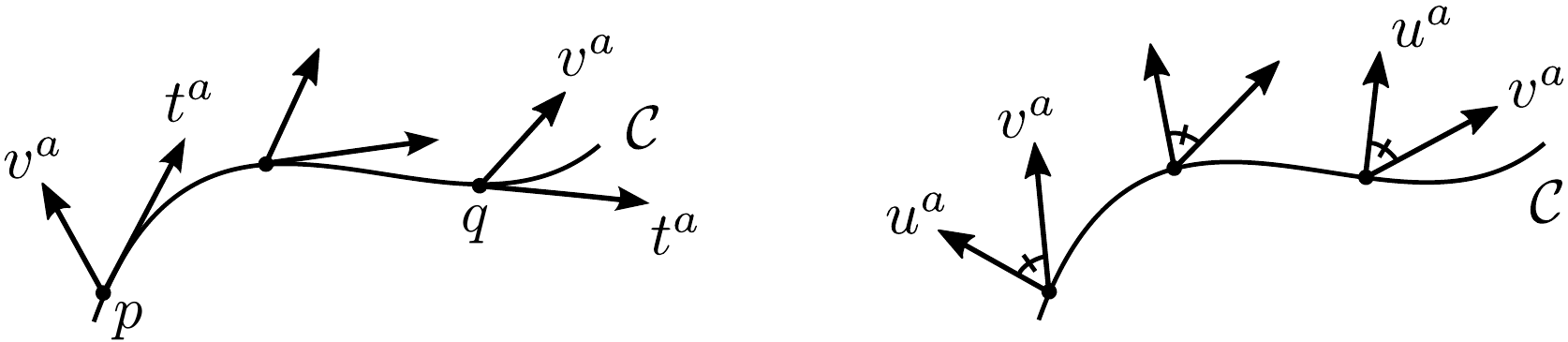}
		\caption{\!\!Illustration of the notion of parallel transport. \textsl{Left panel:} a vector $v^a$ is parallelly transported along a curve $\mathcal{C}$ with tangent vector $t^a$. \textsl{Right panel:} for a metric-compatible connection, the scalar product $g_{ab} u^a v^b$ between two vectors $u^a$ and $v^a$ that are parallelly transported along a curve $\mathcal{C}$ is conserved.}
		\label{f:transport}
	\end{center}
	\vspace{-0.2cm}
\end{figure}

\subsubsection*{Parallel transport}

A connection $\n_a$ can be used to compare two vectors that belong to different tangent spaces, thanks to the notion of \emph{parallel transport} of a vector along a curve. \!More precisely, a vector $v^a$ is said to be parallelly transported along a curve $\mathcal{C}$ with tangent vector $t^a$ if, and only if,
\be\label{e:transp_paral}
	t^a \n_a v^b = 0 \p
\ee
This is the generalization to the case of a manifold of the notion of
``keeping a vector constant'' in ordinary vector spaces. Using the
expressions \eqref{e:def_vector2} and \eqref{e:der_tensor},
the components of \eref{e:transp_paral} with respect to a coordinate
basis read
\be\label{e:transp_paral_tensor}
	\frac{\ud v^\alpha}{\ud \lambda} + C^\alpha_{\phantom{\alpha}\beta\gamma} \, t^\beta v^\gamma = 0 \p
\ee
This shows that, given a vector $v^a$ at a point $p \in \mathcal{C}$ and a connection $C^\alpha_{\phantom{\alpha}\beta\gamma}$, the operation of parallel transport defines a unique vector $v^a$ at any point $q$ along that curve; see Fig.~\ref{f:transport}. The notion of parallel transport can, naturally, be generalized to a generic tensor of type $(r,s)$.

\subsubsection*{Levi-Civita connection}

Let us consider a curve $\mathcal{C}$ with tangent vector $t^a$, as
well as two vector fields $u^a$ and $v^a$ that fulfill the equation of
parallel transport \eqref{e:transp_paral}. Given a metric $g_{ab}$, it
is natural to request that the scalar product $g_{ab} \, u^a v^b$ is
conserved by the parallel transport associated with the
connection $\n_a$:
\be\label{e:transp_prod_scal}
  t^c \n_c \bigl( g_{ab} \, u^a v^b \bigr) = 0 \p
\ee
In particular, this would imply that the squared norms $g_{ab} u^a u^b$ and $g_{ab} v^a v^b$, as well as the angle between $u^a$ and $v^a$ are also preserved by parallel transport along $\mathcal{C}$; see Fig.~\ref{f:transport}. The requirement that the property \eqref{e:transp_prod_scal} holds true for all curves and for all vector fields implies
\be\label{e:compat_metric}
	\n_{c\,} g_{ab} = 0 \p
\ee
A covariant derivative $\n_a$ that satisfies this condition is said to
be \emph{compatible} with the metric $g_{ab}$.

Interestingly, the fundamental theorem of Riemannian geometry stipulates that given a
metric $g_{ab}$, there exists a unique connection $\n_a$ compatible
with that metric. The connection coefficients
$C^\gamma_{\phantom{\gamma}\alpha\beta}$ are then denoted
$\Gamma^\gamma_{\phantom{\gamma}\alpha\beta}$, and referred to as the
\emph{Christoffel symbols}; they read
\be\label{e:def_Christoffel}
\Gamma^\gamma_{\phantom{\gamma}\alpha\beta} = \frac{1}{2} \,
g^{\gamma\delta} \left( \partial_\alpha g_{\delta\beta}
  + \partial_\beta g_{\alpha\delta} - \partial_\delta g_{\alpha\beta}
\right) .  \ee
Thanks to the condition \eqref{e:no_torsion} of absence of torsion,
the Christoffel symbols are symmetric under exchange of the lower two
indices: $\Gamma^\gamma_{\phantom{\gamma}\alpha\beta} =
\Gamma^\gamma_{\phantom{\gamma}\beta\alpha}$. Such a connection is
called a Riemannian connection, or \emph{Levi-Civita connection}. It
is the connection used to formulate the general theory of relativity.

\subsubsection*{Geodesics}\label{ss:geodesics}

Intuitively, a geodesic is a curve whose curvature is ``as small as
possible,'' namely the straightest path possible between two points in
a curved space. Mathematically, given a metric $g_{ab}$ and the
associated Levi-Civita connection $\n_a$, a \emph{geodesic} is a curve
whose tangent vector is parallelly transported along itself,
i.e., a curve such that
\be\label{e:geodesic}
	t^a \n_a t^b = 0 \p
\ee
Geodesics are the natural generalization to curved spaces (and spacetimes) of the straight lines of ordinary Euclidean geometry.

In order to develop some intuition about geodesics, we introduce a coordinate system $(x^\alpha)$ and consider the components of \eref{e:geodesic} with respect to the associated coordinate basis ${(\partial_\alpha)}^a$. If $x^\alpha = X^\alpha(\lambda)$ is a parameterization of the geodesic, then the components of the tangent vector $t^a$ are given by $t^\alpha = \ud X^\alpha / \ud \lambda \equiv \dot{X}^\alpha$. Applying the general formula \eqref{e:transp_paral_tensor} to the case where $v^a$ coincides with $t^a$, we find (for all $0 \leqslant \alpha \leqslant 3$)
\be\label{e:geodesic2}
	\ddot{X}^\alpha + \Gamma^\alpha_{\phantom{\alpha}\beta\gamma} \, \dot{X}^\beta \dot{X}^\gamma = 0 \p
\ee
This is a system of four nonlinear, coupled, second-order, ordinary differential equations for the four functions $X^\alpha(\lambda)$. Given initial conditions $X^\alpha(\lambda_0)$ and $\dot{X}^\alpha(\lambda_0)$, Cauchy's theorem implies that this system has a unique solution. Thus, for all $p \in \mathcal{M}$, there is a unique geodesic going through $p$ with a given spacetime direction.

\subsection{Worldlines and proper time}\label{ss:proper_time}

In general relativity, the path of a massless particle (or photon) in spacetime is a \emph{null geodesic}, i.e., a curve whose tangent vector is lightlike everywhere, and which obeys the geodesic equation \eqref{e:geodesic}. Just like in special relativity, the paths of all photons that go through a given event $p \in \mathcal{M}$ define a local lightcone at $p$, an intrinsic structure in spacetime.

\begin{figure}
	\begin{center}
		\includegraphics[width=12cm]{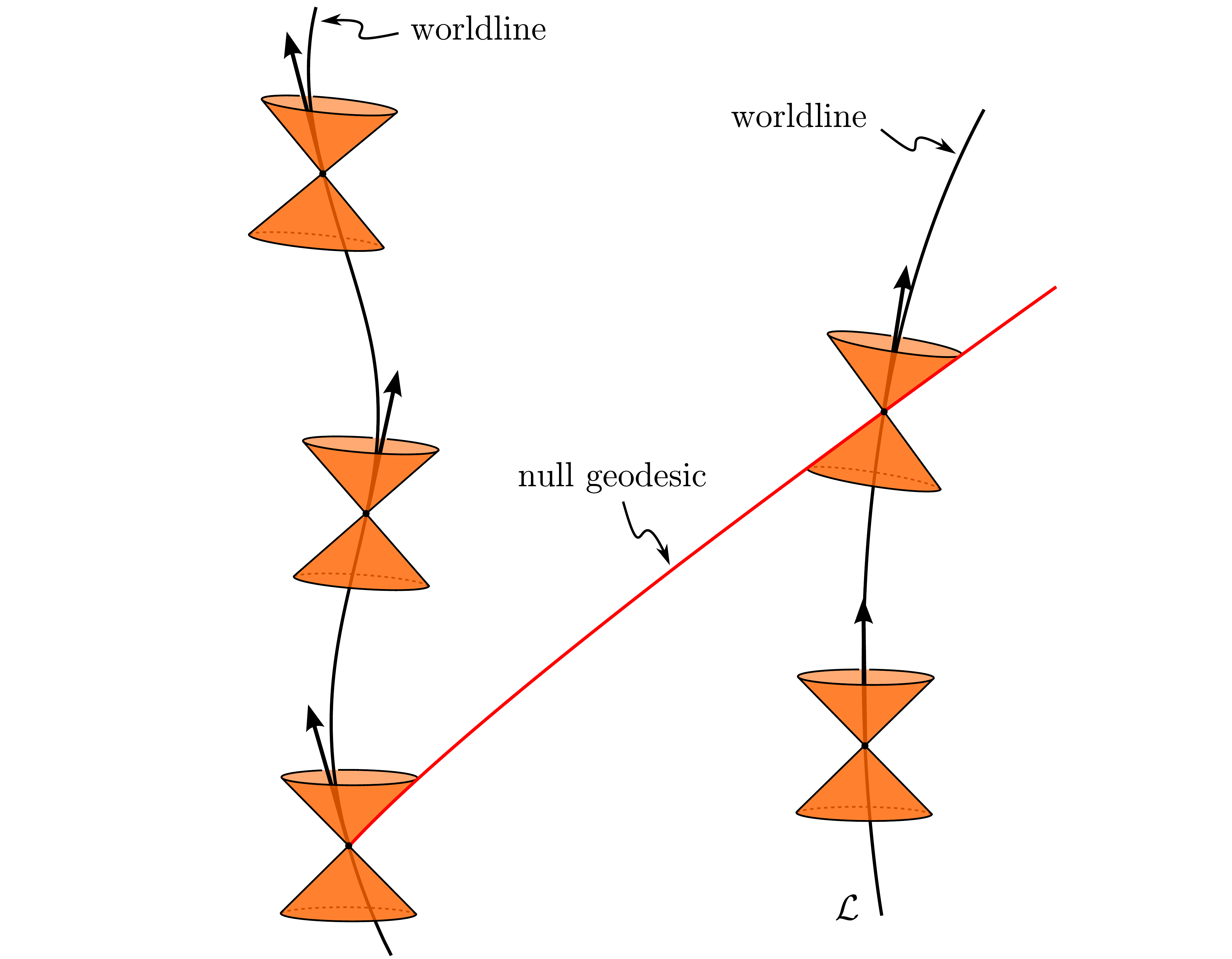}
		\caption{The tangent vector to a wordline $\mathcal{L}$ always lies within the locally-defined lightcone. Note that, contrary to \fref{f:lightcone}, lightcones can appear tilted in the curved spacetimes of general relativity. This is because in presence of a gravitational field there exists no global inertial frame. (One spatial dimension is not shown.)}
		\label{f:worldline}
	\end{center}
\end{figure}

On the other hand, the path of a massive particle (or physical body) in spacetime is a \emph{worldline}, namely a curve whose tangent vector is timelike everywhere. The tangent vector to a worldline $\mathcal{L}$ must necessarily lie within the local lightcone for all $p \in \mathcal{L}$, as depicted in \fref{f:worldline}. This is the geometrical translation of the fact that massive particles cannot travel faster than light.

\vspace{0.3cm}

\begin{figure}
	\begin{center}
		\includegraphics[width=12cm]{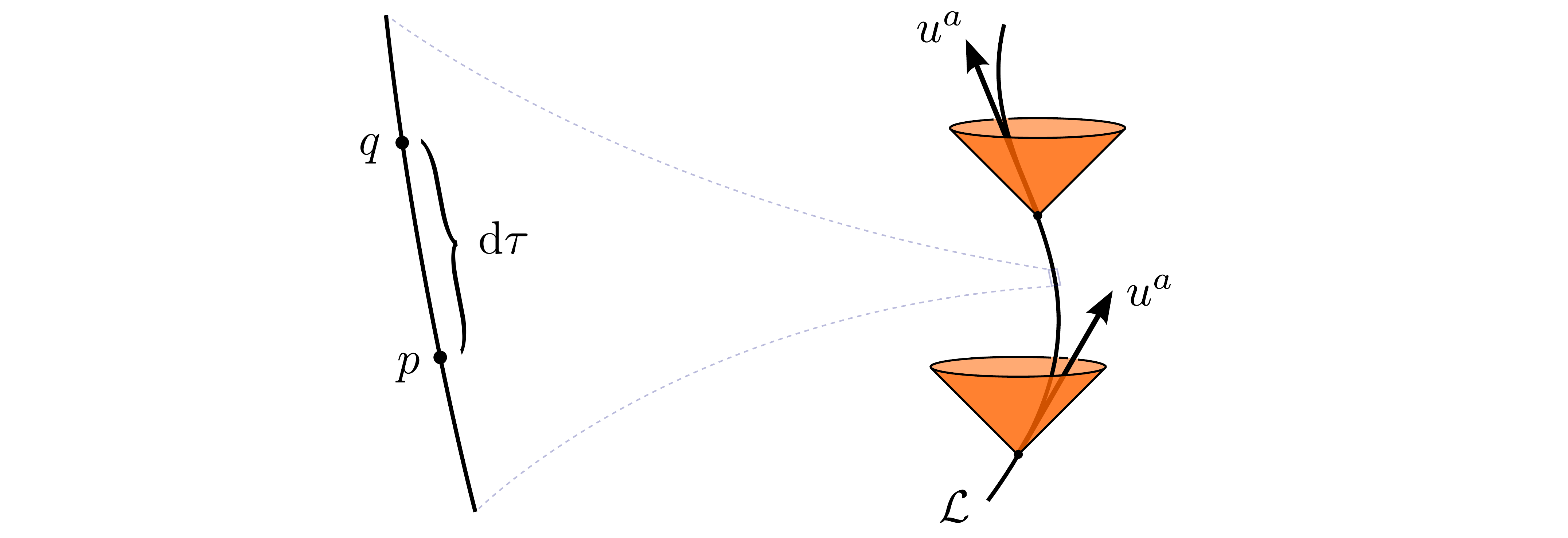}
		\caption{The worldline $\mathcal{L}$ of a particle can be parameterized by the proper time $\tau$ elapsed along that curve. The associated tangent vector is the four-velocity $u^a$ of the particle.}
		\label{f:proper_time}
	\end{center}
	\vspace{-0.8cm}
\end{figure}

In general relativity, a key interpretation of the metric has to do with the measure of time along the worldline $\mathcal{L}$ of a particle. Let $p$ and $q$ denote two nearby events along $\mathcal{L}$. Since the tangent vector to this curve is timelike, the interval \eqref{e:ds2_GR} between $p$ and $q$ is negative. Then, the \emph{proper time} elapsed between those events along the worldline $\mathcal{L}$ is the number (see \fref{f:proper_time})
\be\label{e:def_proper_time}
	\ud\tau \equiv \bigl( - \ud s^2 \big)^{1/2} .
\ee
This is the physical time that an ideal clock moving with the particle would measure between $p$ and $q$. Given a coordinate system $(x^\alpha)$, if $x^\alpha = X^\alpha(\lambda)$ is a parameterization of $\mathcal{L}$, then the definition \eqref{e:def_proper_time} can be written as
\be\label{e:proper_time}
	\ud\tau = \bigl( - g_{\alpha\beta} \dot{X}^\alpha \dot{X}^\beta \big)^{1/2} \, \ud \lambda \p
\ee

The proper time elapsed along the worldline of a particle yields a natural parameterization of that curve. The tangent vector $t^a$ associated with $\lambda = \tau$ is the \emph{four-velocity} $u^a\!$ of the particle, whose components with respect to the coordinate basis vectors ${(\partial_\alpha)}^a$ read
\be\label{e:def_4vel}
	u^\alpha = \frac{\ud x^\alpha}{\ud \tau} = \frac{\dot{X}^\alpha}{(- g_{\beta\gamma} \dot{X}^\beta \dot{X}^\gamma)^{1/2}} \p
\ee
Equation \eqref{e:proper_time} implies that the four-velocity is a timelike vector whose norm squared is constant and equal to $g_{ab} u^a u^b = -1$.

An \emph{observer} is modelled as a worldline in spacetime with a four-velocity $u^a$. The worldline of a \emph{freely falling} observer is a timelike geodesic, a curve whose tangent vector $u^a$ obeys the geodesic equation \eqref{e:geodesic}. Such worldlines have the property of maximazing (locally) the proper time elapsed between any two events $p$ and $q$. Indeed, using the expression \eqref{e:proper_time}, it can be shown that the condition $\delta \int_p^q \ud \tau = 0$ is equivalent to the geodesic equation \eqref{e:geodesic2}.

\section{Spacetime Curvature and Matter}\label{s:Einstein}

Having introduced the geometrical setting of the theory of general relativity, we move on to the mathematical description of spacetime curvature and its generation by the energy and momentum of matter.

\subsection{Riemann curvature tensor}\label{ss:Riemann}

As we have seen in \sref{ss:cov_deriv}, two covariant derivatives acting on a scalar field commute [condition \eqref{e:no_torsion} of absence of torsion]. However, this property does not hold true for tensor fields of higher ranks. In particular, for a vector field $v^a$ and a field of dual vectors $\omega_a$, we have
\begin{subequations}\label{e:def_Riemann}
	\begin{align}
		\n_a \n_b \, v^c - \n_b \n_a \, v^c &= R^c_{\phantom{c}dab} \, v^d \c \\
		\n_a \n_b \, \omega_c - \n_b \n_a \, \omega_c &= - R^d_{\phantom{d}abc} \, \omega_d \c
	\end{align}
\end{subequations}		
where $R^a_{\phantom{a}bcd}$ is a tensor of type $(1,3)$ that is known as the \emph{Riemann curvature tensor}. Its tensorial nature is obvious from Eqs.~\eqref{e:def_Riemann}, because the covariant derivative of a tensor is itself a
tensor. By combining the formulas \eqref{e:def_Riemann} with \eqref{e:der_tensor}, it can be shown that the components of the Riemann curvature tensor with respect to a given coordinate basis read
\be\label{e:formula_Riemann}
	R^\alpha_{\phantom{\alpha}\beta\mu\nu} = \partial_\mu \Gamma^\alpha_{\phantom{\alpha}\beta\nu} - \partial_\nu \Gamma^\alpha_{\phantom{\alpha}\beta\mu} + \Gamma^\alpha_{\phantom{\alpha}\sigma\mu} \Gamma^\sigma_{\phantom{\sigma}\beta\nu} - \Gamma^\alpha_{\phantom{\alpha}\sigma\nu} \Gamma^\sigma_{\phantom{\sigma}\beta\mu} \p
\ee
The Riemann tensor can be given several interpretations. In particular, it is related to (i) the failure of a vector to come back to itself after having been parallelly transported along a small loop, and (ii) the relative acceleration of two nearby geodesics.

\subsubsection*{Parallel transport and curvature}

\begin{figure}
	\begin{center}
		\includegraphics[width=12cm]{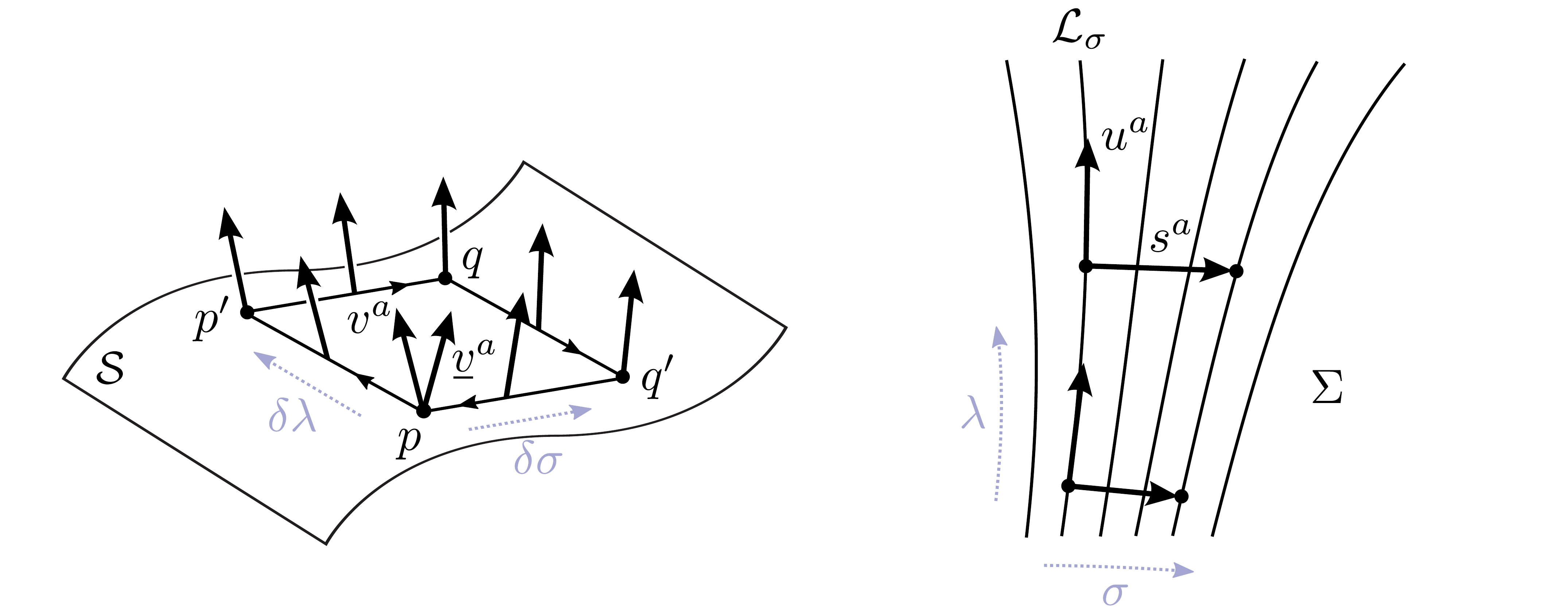}
		\caption{Two illustrations of the effects of curvature. \textsl{Left panel:} A vector that undergoes parallel transport along a small closed curve does not come back to itself. \textsl{Right panel:} Nearby geodesics ``accelerate'' relative to one another.}
		\label{f:curvature}
	\end{center}
\end{figure}

Using a surface $\mathcal{S}$, let us construct an infinitesimal closed curve $\mathcal{C}$ (a loop) at a point $p$. If $(x^\alpha) = (\lambda,\sigma)$ is a coordinate system on $\mathcal{S}$, let $p$, $p'$, $q$ and $q'$ be the points of coordinates $(0,0)$, $(\delta\lambda,0)$, $(\delta\lambda,\delta\sigma)$ and $(0,\delta\sigma)$; see \fref{f:curvature}. Let $\delta\lambda \, u^a$ and $\delta\sigma \, w^a$ denote the vectors that connect $p$ to the points $p'$ and $q'$. Now, if $\underline{v}^a$ denotes the result of the parallel transport of a vector $v^a \in \mathcal{T}_p$ along the loop $\mathcal{C}$, then the difference $\delta^{(2)} v^a \equiv \underline{v}^a - v^a$ is given by
\be\label{e:curv_transport}
	\lim_{{\delta \lambda \to 0} \atop {\delta \sigma \to 0}} \, \frac{\delta^{(2)} v^a}{\delta \lambda \, \delta \sigma} = R^a_{\phantom{a}bcd} \, v^b u^c w^d \, .
\ee
Thus, a vector that is parallelly transported along a small closed curve does not come back to itself. Equivalently, the result of the parallel transport of a vector between two points ($q$ and $p$ in \fref{f:curvature}) depends on the path chosen.

\subsubsection*{Equation of geodesic deviation}

Let us consider a family of geodesics $(\mathcal{L}_\sigma)_{\sigma \in I}$, where the parameter $\sigma$ ranges in an interval $I \subset \mathbb{R}$. Each curve $\mathcal{L}_\sigma$ is a geodesic parameterized by $\lambda \in \mathbb{R}$. Let $\Sigma \subset \mathcal{M}$ be the two-dimensional submanifold generated by these curves, and $(x^\alpha) = (\lambda,\sigma)$ a coordinate system on $\Sigma$. The vector field $u^a \equiv {(\partial_\lambda)}^a$ is tangent to each curve, while $s^a \equiv {(\partial_\sigma)}^a$ can be interpreted as the \emph{separation vector} between $\mathcal{L}_\sigma$ and a nearby geodesic (see \fref{f:curvature}).

Intuitively, the covariant derivative $\dot{s}^a \equiv u^b \nabla_b s^a$ of the separation vector along $\mathcal{L}_\sigma$ is the ``relative velocity'' of two nearby geodesics. Therefore, the covariant derivative $\ddot{s}^a \equiv u^c \nabla_c (u^b \nabla_b s^a)$ of that vector along $\mathcal{L}_\sigma$ can be interpreted as their ``relative acceleration.'' It can be shown that the evolution of the relative velocity is controlled by the curvature tensor through the \emph{equation of geodesic deviation}
\be\label{e:geo_dev}
	\ddot{s}^a = R^a_{\phantom{a}bcd} \, u^b u^c s^d \p
\ee
This equation holds to linear order in the separation vector $s^a$. In absence of curvature, two neighboring geodesics that are initially parallel ($\dot{s}^a = 0$) will remain parallel. If the curvature tensor does not vanish, however, two such curves will get closer or further apart. This is the case, for instance, of lines of longitude at the surface of a sphere, namely geodesics that are orthogonal to the  equator but that meet at the poles. According to \eref{e:geo_dev}, the ``relative acceleration'' between two nearby geodesics vanishes for all families of geodesics if, and only if, $R^a_{\phantom{a}bcd} = 0$.

\subsubsection*{Properties of the Riemann tensor}

A spacetime whose curvature tensor vanishes over the entire manifold
$\mathcal{M}$ is said to be \emph{flat}. Using the definition
\eqref{e:def_Riemann}, or any of the above two properties, it can be
shown that the only flat spacetime is that of special relativity,
i.e. 
\be
  \label{e:flat_Riemann}
  R^a_{\phantom{a}bcd} = 0 \iff g_{ab} = \eta_{ab} \p
\ee
Importantly, the Riemann curvature tensor fulfills some algebraic
identities: (i) it is antisymmetric with respect to the first and the
last pairs of indices, and (ii) it has a cyclic symmetry with respect
to the last three indices: 
\begin{subequations}\label{e:sym-cycl}
	\begin{align}
		R_{abcd} &= - R_{bacd} = - R_{abdc} \c \label{e:sym} \\
		R^a_{\phantom{a}bcd} &+ R^a_{\phantom{a}dbc} + R^a_{\phantom{a}cdb} = 0 \p \label{e:cycl}
	\end{align}
\end{subequations}
These two properties can be combined to establish that the Riemann
tensor is also symmetric under exchange of the first and last pairs of
indices, that is $R_{abcd} = R_{cdab}$. As a consequence, only 20 out of
the $4^4 = 256$ components of the curvature tensor are linearly
independent. Moreover, the tensor $R^a_{\phantom{a}bcd}$ obeys the \emph{Bianchi
  identity}, an important differential identity that reads 
\be\label{e:Bianchi}
	\n_e R^a_{\phantom{a}bcd} + \n_d R^a_{\phantom{a}bec} + \n_c R^a_{\phantom{a}bde} = 0 \p
\ee

\subsection{Ricci tensor and scalar curvature}\label{ss:Ricci}

From the Riemann curvature tensor it is possible to define other, lower-rank
tensors, such as the \emph{Ricci tensor}
\be\label{e:def_Ricci}
	R_{ab} \equiv g^{cd} R_{cadb} = R^c_{\ acb} \c
\ee
which is symmetric. The Ricci tensor is the only nontrivial rank-two
tensor that can be obtained by contracting a pair of indices of the
Riemann tensor. Indeed, because of the symmetry properties
\eqref{e:sym-cycl}, other contractions yield $\pm R_{ab}$ or vanish
identically. The trace of the Ricci tensor, 
\be\label{e:def_Ricci_scal}
	R \equiv g^{ab} R_{ab} = R^{ab}_{\phantom{ab}ab} \c
\ee
is called the \emph{scalar curvature}. It is the only nonzero scalar
field that can be constructed by contracting two pairs of indices of the Riemann tensor.

Finally, when contracted over the first and last indices, as well as on the
second and third indices, the Bianchi identity \eqref{e:Bianchi}
yields
\be\label{e:divfree}
	\nabla^a \Bigl( R_{ab} - \frac{1}{2} R \, g_{ab} \Bigr) = 0 \c
\ee
a relation that is known as the contracted Bianchi identity. The
divergence-free tensor $G_{ab} \equiv R_{ab} - \frac{1}{2} R \,
g_{ab}$ is the \emph{Einstein tensor}. As will be discussed
in \sref{ss:Einstein_eq}, this tensor plays a central role in
the Einstein equation of general relativity. Notice that none of the
conditions $R=0$, $R_{ab} = 0$, or $G_{ab} = 0$ necessarily implies that
spacetime is flat.

\subsection{Energy-momentum tensor}\label{ss:Tmunu}

In Newtonian gravity, the gravitational potential is generated by the
distribution of the matter mass density. In general relativity, all
types of matter and radiation produce a gravitational field through
their \emph{energy-momentum tensor} (or stress-energy tensor), a
symmetric rank-two tensor $T_{ab}$ that has the dimensions of an
energy density.

A formal definition of that tensor in terms of a Lagrangian
formulation can be used to prove that $T_{ab}$ must be divergence-free:
\be
  \label{e:div_T}
  \nabla^a T_{ab} = 0 \p
\ee
This equation expresses the law of local conservation of energy
and momentum. Indeed, according to an observer with a four-velocity
$u^a$, the energy density in the matter fields is given by the scalar 
\be\label{e:energy}
	\varepsilon = T_{ab} \, u^a u^b \p
\ee
Moreover, according to that same observer, the density of linear
momentum along the spatial direction $e_i^a$ (such that $g_{ab} e_i^a u^b = 0$ and $g_{ab} e_i^a e_j^b = \delta_{ij}$) and the flux of energy along that same direction are both given by the scalar
\be\label{e:momentum}
	P_i = - \, T_{ab} \, e_i^a u^b \quad (1 \leqslant i \leqslant 3) \p
\ee

For a given type of matter (e.g. dust, electromagnetic field, scalar
field), the energy-momentum tensor can easily be derived from the
corresponding Lagrangian. For instance, for a \emph{perfect fluid}
with a four-velocity field $u^a$, an energy density $\varepsilon$
and a pressure $P$, the energy-momentum tensor reads
\be\label{e:def_perfect_fluid}
	T_{ab} = (\varepsilon + P) \, u_a u_b + P \, g_{ab} \p
\ee
Note that the metric $g_{ab}$ does, in general, enter explicitly the
expression for the energy-momentum tensor. Equation \eqref{e:def_perfect_fluid} is, of course, compatible with the interpretations given to the quantities \eqref{e:energy} and \eqref{e:momentum}.

\subsection{Einstein's equation}\label{ss:Einstein_eq}

We have introduced all of the concepts required to formulate the field
equation of general relativity. As we shall see, Einstein's equation
relates part of the curvature of spacetime (the Einstein tensor
$G_{ab}$) to its matter content (the energy-momentum tensor $T_{ab}$)
and it reduces to Poisson's equation \eqref{e:Poisson} of Newtonian
gravity in the nonrelativistic limit where $c^{-1} \to 0$. In what
follows, we motivate Einstein's equation along the lines of Ref.~\refcite{Wal}.

In Newtonian gravity, the equation of motion of a particle with position $\vec{x}$ reads $\ddot{\vec{x}} = - \vec{\nabla} \Phi$, where $\Phi$ is the Newtonian potential. If $\vec{s} \equiv \vec{x}_1 - \vec{x}_2$ denotes the relative position of two nearby particles, then a Taylor expansion shows that their relative acceleration is given, to leading order, by
\be
	\ddot{\vec{s}} = - (\vec{s} \cdot \vec{\nabla}) \vec{\nabla} \Phi \, .
\ee
This equation is, quite clearly, analogous to the equation of geodesic deviation \eqref{e:geo_dev}. It suggests the following analogy between the Riemann curvature tensor and the Newtonian tidal field:
\be\label{e:R_ddPhi}
	R^a_{\phantom{a}bcd} u^c u^d \, \longleftrightarrow \, \partial^a \partial_b \Phi \, .
\ee
Moreover, still in Newtonian gravity, the trace $\nabla^2 \Phi = \partial^a \partial_a \Phi$ of the Newtonian tidal field is related to the mass density $\rho$ of matter through Poisson's equation \eqref{e:Poisson}. On the other hand, as discussed in Sec.~\ref{ss:Tmunu} above, in general relativity all ``matter'' fields are described by an energy-momentum tensor $T_{ab}$ such that
\be\label{e:T_rho}
	T_{ab} \, u^a u^b \, \longleftrightarrow \, \rho \, .
\ee

Thus, by combining the analogies \eqref{e:R_ddPhi} and \eqref{e:T_rho} with Poisson's equation \eqref{e:Poisson}, we are led to postulate an equation of the form $R_{ab} u^a u^b = 4 \pi G \, T_{ab} u^a u^b$. Because this equation must hold true for all observers with four-velocity $u^a$, these physical arguments suggest the field equation
\be
	R_{ab} \stackrel{?}{=} 4 \pi G \, T_{ab} \, .
\ee
Unfortunately, this relationship is flawed. Since the energy-momentum tensor is conserved, $\nabla^a T_{ab} = 0$, the proportionality of $R_{ab}$ and $T_{ab}$ together with the contracted Bianchi identity \eqref{e:divfree} would imply that $\n_a R = 0$, i.e., that $R = g^{ab} R_{ab}$ and therefore $T \equiv g^{ab} T_{ab}$ are constant throughout spacetime. This restriction on the energy contents of the Universe is too strong.

Nevertheless, this difficulty suggests a natural resolution. To avoid the conflict between the conservation of energy and momentum on one hand, and the contracted Bianchi identity on the other hand, one simply has to postulate \emph{Einstein's equation}\footnote{The most general formulation of Einstein's equation involves the additional term $\Lambda \, g_{ab}$ in the left-hand side of \eqref{e:Einstein}, where $\Lambda$ is the \emph{cosmological constant}, measured to the value $\Lambda \simeq 10^{-52}~\text{m}^{-2}$. Outside of cosmology, this additional term can safely be neglected.}
\be
  \label{e:Einstein}
  R_{ab} - \frac{1}{2} R\, g_{ab} = 8 \pi G \, T_{ab} \p
\ee
Indeed, if \eref{e:Einstein} is satisfied, then the local conservation of energy and momentum becomes a consequence of the (contracted) Bianchi identity. Moreover, the analogies \eqref{e:R_ddPhi} and \eqref{e:T_rho} that have motivated this field equation are unaffected. Indeed, taking the trace of \eref{e:Einstein}, one obtains $R = - 8 \pi G \, T$, such that Einstein's equation can be rewritten in the equivalent form
\be\label{e:Einstein_alt}
	R_{ab} = 8 \pi G \left( T_{ab} - \frac{1}{2} \, T g_{ab} \right) .
\ee
In the nonrelativistic limit, the energy density dominates all the other contributions to the energy-momentum tensor, such that $T_{ab} u^a u^b \simeq - T \simeq \rho$. Hence, \eref{e:Einstein_alt} still implies the relation $R_{ab} u^a u^b = 4 \pi G \, T_{ab} u^a u^b$ for weak gravitational fields.

The coupling constant in the right-hand sides of Eqs.~\eqref{e:Einstein}--\eqref{e:Einstein_alt} ensures that Einstein's equation reduces to Poisson's equation \eqref{e:Poisson} in the appropriate limit. Restoring powers of $c^{-1}$, its numerical value,
\be
\frac{8 \pi G}{c^4} \simeq 2 \times 10^{-43}~\frac{\text{m}^{-2}}{\text{J}
  \cdot \text{m}^{-3}} \c 
\ee
shows that a large amount of energy density is required in order to
produce spacetime curvature, which is homogeneous to an inverse length
squared.

Once a coordinate system has been chosen, Einstein's equation \eqref{e:Einstein} becomes equivalent to a set of ten second-order, non-linear partial differential equations for the unknown metric components $g_{\alpha\beta}(x)$. However, because of the contracted Bianchi identity \eqref{e:divfree}, only six of these ten partial differential equations are independent, and the freedom in the choice of coordinates can be used to specify four out of the ten metric components $g_{\alpha\beta}(x)$. Therefore, one is left with six equations for six unknowns.

\section{Definition of Gravitational Waves}\label{s:def_gw}

In the previous sections, we have introduced the general theory of relativity. At long last, we are in a position to define the concept of gravitational wave, whose existence was first predicted in 1916 by Einstein himself.\cite{Ei.16}

\subsection{Linearized Einstein equation}\label{ss:linear}

Far away from compact objects (black holes and neutrons stars),
gravitation is ``weak'' in the sense that the spacetime geometry is
nearly flat. Therefore, in most astrophysical situations, the physical
metric $g_{ab}$ is ``close'' to the Minkowski metric $\eta_{ab}$ of
special relativity, in the sense that
\be\label{e:def_h}
	g_{ab} = \eta_{ab} + h_{ab} \c
\ee
with $h_{ab}$ a ``small'' metric perturbation. Since there is no
natural positive-definite metric on spacetime, there is no natural
norm by which ``smallness'' of tensors can be measured. However, we
may require that, with respect to an inertial coordinate system
of $\eta_{ab}$, for which $\eta_{\alpha\beta} = \text{diag} \,
(-1,+1,+1,+1)$, the components $h_{\alpha\beta}$ of $h_{ab}$ obey
\be\label{e:h<<1}
	|h_{\alpha\beta}| \ll 1 \p
\ee
Then, by substituting for \eref{e:def_h} in the identity $g^{ab} g_{bc} = \delta^a_{\phantom{a}c}$, and making use of $\eta^{ab} \eta_{bc} = \delta^a_{\phantom{a}c}$, the inverse metric $g^{ab}$ is also found to be ``close'' to the inverse Minkowski metric $\eta^{ab}$, in the sense that
\be\label{e:def_h-1}
	g^{ab} = \eta^{ab} - h^{ab} \c
\ee
where we introduced the notation $h^{ab} \equiv \eta^{ac}
\eta^{bd} h_{cd}$ and neglected all terms $\calO(h^2)$. Thereafter, we
will work to \emph{linear order} in $h_{ab}$ and omit all remainders $o(h)$. All indices will thus be
``lowered'' and ``raised'' by using the flat metric $\eta_{ab}$ and
its inverse $\eta^{ab}$.

We may then proceed to linearize the Einstein equation
\eqref{e:Einstein} with respect to the metric perturbation
$h_{ab}$. Substituting for Eqs.~\eqref{e:def_h} and \eqref{e:def_h-1}
into the expression \eqref{e:formula_Riemann} for the Riemann tensor, and
using the explicit formula \eqref{e:def_Christoffel} for the
Christoffel symbols, we find
\be\label{e:linear_Riemann}
	R_{abcd} = - \partial_c \partial_{[a} h_{b]d} + \partial_d \partial_{[a} h_{b]c} \c
\ee
where $\partial_a$ is the ordinary derivative associated with the global inertial coordinates of $\eta_{ab}$, and square brackets are used to denote an antisymetrization over a pair of indices, e.g., $T_{[ab]} \equiv \frac{1}{2} (T_{ab} - T_{ba})$. Using the definitions \eqref{e:def_Ricci}--\eqref{e:def_Ricci_scal} of the Ricci tensor and scalar curvature, the Einstein tensor $G_{ab}$ can be linearized as well. Einstein's equation \eqref{e:Einstein} then reduces to
\be\label{e:box_hmunu}
	\Box \bar{h}_{ab} - 2 \partial_{(a} V_{b)} + \eta_{ab} \, \partial^c V_c = - 16 \pi G \, T_{ab} \c
\ee
where $\Box \equiv \eta^{cd} \partial_c \partial_d$ is the usual flat-space wave operator, or d'Alembertian, and parenthesis are used to denote a symetrization over a pair of indices, e.g. $T_{(ab)} \equiv \frac{1}{2} (T_{ab} + T_{ba})$. Moreover, we introduced the notation $V_a \equiv \partial^c \bar{h}_{ac}$ for the divergence of the trace-reversed metric perturbation
\be\label{e:def_hbar}
	\bar{h}_{ab} \equiv h_{ab} - \frac{1}{2} h \, \eta_{ab} \c
\ee
with $h \equiv \eta^{ab} h_{ab}$. Computing the trace of $\bar{h}_{ab}$ yields $\bar{h} = - h$, such that \eqref{e:def_hbar} can easily be inverted to give $h_{ab} = \bar{h}_{ab} - \frac{1}{2} \bar{h} \, \eta_{ab}$. Equation \eqref{e:box_hmunu} takes on a slightly more complicated form in terms of the metric perturbation $h_{ab}$.

\subsection{Lorenz gauge condition}

Interestingly, the form \eqref{e:def_h}--\eqref{e:h<<1} of the metric does, by no means, uniquely specify the perturbation $h_{ab}$. The freedom to perform ``infinitesimal'' coordinate transformations $x^\alpha \to x^\alpha - \xi^\alpha(x)$ that preserve the form \eqref{e:def_h}--\eqref{e:h<<1} of the metric gives rise, in the linearized theory, to an invariance under \emph{gauge transformations} of the form
\be\label{e:change_h}
	h_{ab} \to h_{ab} + 2 \partial_{(a} \xi_{b)} \p
\ee
The arbitrary vector field $\xi^a$ is known as the \emph{generator} of the gauge transformation \eqref{e:change_h}. It can be checked that the linearized Riemann tensor \eqref{e:linear_Riemann} is invariant under such a transformation. This gauge freedom is analogous to that of ordinary electromagnetism in flat spacetime, where the Faraday tensor $F_{ab} = \partial_{[a} A_{b]}$ is invariant under a gauge transformation $A_a \to A_a + \partial_a \chi$ of the vector potential $A_a$ generated by an arbitrary function $\chi$.

The gauge freedom of linearized gravitation can be used to simplify the linearized Einstein equation \eqref{e:box_hmunu}. In particular, one can always find a gauge in which the divergence $V_a$ vanishes, i.e., such that
\be\label{e:Lorenz_gauge}
	\partial^c \bar{h}_{ac} = 0 \p
\ee
Notice that $V_a \to V_a + \Box \xi_a$ under the gauge transformation \eqref{e:change_h}. Therefore, starting from a gauge where $V_a \neq 0$, one moves to a gauge obeying \eqref{e:Lorenz_gauge} by applying a gauge transformation with a generator solution of $\Box \xi_a = - V_a$.

By analogy with the gauge condition $\partial^a \! A_a = 0$ of electromagnetism, Eq. \eqref{e:Lorenz_gauge} is known as the \emph{Lorenz gauge} condition, or \emph{harmonic gauge} condition. In the Lorenz gauge, the linearized Einstein equation \eqref{e:box_hmunu} reduces to
\be
  \label{e:box_h}
  \Box \bar{h}_{ab} = - 16 \pi G \, T_{ab} \p
\ee
Thus, $\bar{h}_{ab}$ represents a quantity that propagates as a wave at the vacuum speed of light, on a flat Minkowski background, and which is sourced by the energy-momentum tensor $T_{ab}$ of matter; in other words, $\bar{h}_{ab}$ is a \emph{gravitational wave}. Given a matter source, the solution of the wave equation \eqref{e:box_h} for each component $\bar{h}_{\alpha\beta}$ of $\bar{h}_{ab}$ is a standard problem in physics---familiar, for example, from the theory of electromagnetic waves.

The linearized Einstein equation \eqref{e:box_h} is reminiscent of the Lorenz-gauge Maxwell equation $\Box A_a = - \mu_0 \, j_a$, with $\mu_0$ the vacuum permeability and $j^a$ the current density. Just like the gauge condition $\partial^a \! A_a = 0$ implies the local conservation of the electric charge, $\partial^a \! j_a = 0$, the harmonic gauge condition \eqref{e:Lorenz_gauge} implies the local conservation of energy and momentum in linearized gravity, $\partial^a T_{ab} = 0$.

\vspace{0.4cm}
\begin{table}[h!]
	\tbl{The gauge freedom of linearized gravitation is analogous to that of ordinary electromagnetism in flat spacetime. \vspace{0.1cm}} 
	{\begin{tabular}{@{}lcc@{}}
		\toprule
							& Electromagnetism & Linearized gravity \\
		\midrule
		Generator			& $\chi$							& $\xi_a$ \\ \noalign{\medskip}
		Potential			& $A_a$ 							& $h_{ab}$ \\ \noalign{\medskip}
		Gauge transfo.		& $A_a \to A_a +\partial_a \chi$	& $h_{ab} \to h_{ab} + 2 \partial_{(a} \xi_{b)}$ \\ \noalign{\medskip}
		Gauge invariant		& $F_{ab} = \partial_{[a} A_{b]}$	& $R_{abcd} = - \partial_c \partial_{[a} h_{b]d} + \partial_d \partial_{[a} h_{b]c}$ \\ \noalign{\medskip}
		Lorenz gauge cond.	& $\partial^a \! A_a = 0$			& $\partial^a \bar{h}_{ab} = 0$ \\ \noalign{\medskip}
		Conservation law	& $\partial^a j_a = 0$				& $\partial^a T_{ab} = 0$ \\ \noalign{\medskip}
		Wave equation		& $\Box A_a = - \mu_0 \, j_a$		& $\Box \bar{h}_{ab} = - 16 \pi G \, T_{ab}$ \\
		\bottomrule
	\end{tabular}}
	\label{tab:analogy}
\end{table}

\subsection{Propagation in vacuum}\label{ss:GW_vacuum}

Next, we consider the case of gravitational waves that propagate in vacuum, i.e., we set $T_{ab} = 0$ in Eq.~\eqref{e:box_h}. Together with the harmonic gauge condition \eqref{e:Lorenz_gauge}, the freely propagating waves obey
\be\label{e:box_h0}
	\Box \bar{h}_{ab} = 0 \p
\ee
The general solution of this equation can be written as a linear superposition of monochromatic waves. Thus, we perform the following four-dimensional Fourier decomposition:
\be\label{e:Fourier_h}
	\bar{h}_{ab} (x) = \Re \int A_{ab}(k) \, e^{\ui k_\alpha x^\alpha} \ud^4k \p
\ee
Each Fourier mode has a complex amplitude $A_{ab}(k)$ and is labelled by the components $k_\alpha$ of the wave dual vector $k_a$. Substituting for the ansatz \eqref{e:Fourier_h} into \eref{e:box_h0}, one finds
\be
	\eta^{ab} k_a k_b = 0 \p
\ee
Because gravitational waves propagate at the vacuum speed of light $c$, the wave vector $k^a$ is a null vector (with respect to the Minkowski metric $\eta_{ab}$). On the other hand, the Lorenz gauge condition \eqref{e:Lorenz_gauge} implies that the amplitude tensor is orthogonal to the direction of propagation of the waves:
\be\label{e:Lorenz}
	k^a A_{ab} = 0 \p
\ee

\subsubsection*{Transverse-traceless gauge} 

Notice that the harmonic gauge condition \eqref{e:Lorenz_gauge} does
not, by itself, uniquely specify the metric perturbation
$h_{ab}$. Indeed, any gauge transformation \eqref{e:change_h} whose
generator $\xi^a$ satisfies
\be\label{e:Box_xi}
	\Box \xi_a = 0
\ee
does preserve the gauge condition \eqref{e:Lorenz_gauge}. This is
analogous to the fact that, in ordinary electromagnetism, the Lorenz
gauge condition $\partial^a \! A_a = 0$ does not uniquely fix the
vector potential $A_a$; we have the restricted gauge freedom $A_a \to
A_a + \partial_a \chi$ with $\Box \chi = 0$.

To uniquely specify the metric perturbation, four additional
constraints must be imposed. Let us introduce a unit timelike vector
$u^a$ associated, for instance, with an observer detecting the
gravitational radiation (see Sec.~\ref{s:th_inter}). One can then define a gauge, known
as a \emph{transverse-traceless} (TT) gauge, in which the amplitude
tensor obeys
\begin{subequations}\label{e:TT}
	\begin{align}
		u^a A_{ab} &= 0 \c \label{e:transv} \\
		\eta^{ab} A_{ab} &= 0 \p \label{e:traceless}
	\end{align}
\end{subequations}
Note that the transversality condition \eqref{e:transv} depends upon
the choice of an observer. The tracelessness condition
\eqref{e:traceless} implies $\bar{h} = 0$, such that
$\bar{h}_{\alpha\beta} = h_{\alpha\beta}$ in the TT gauge. Equations
\eqref{e:TT} yield \emph{four} additional constraints that completely
fix the remaining gauge freedom in \eref{e:Box_xi}. Indeed, one of the
four conditions \eqref{e:transv} is redundant with the constraints
\eqref{e:Lorenz}.

\subsubsection*{Polarization states}

Since only eight out of the nine equations \eqref{e:Lorenz} and \eqref{e:TT} are linearly independent, the symmetric $4 \times 4$ amplitude matrix $A_{\alpha\beta}$ has only two independent components left. In the rest frame of the observer, $u^\alpha = (1,0,0,0)$, and for gravitational waves that propagate along the $z$-direction, $k^\alpha = \omega \, (1,0,0,1)$, with $\omega$ the frequency of the wave, as measured by that observer. Equations \eqref{e:Lorenz} and \eqref{e:TT} can then be used to show that the components of the metric perturbation in the TT gauge are given by
\be\label{e:hijTT}
	h_{\alpha\beta}^\text{TT} = \left( 
		\begin{array}{cccc}
			0 & 0 & 0 & 0\\
			0 & h_+ & h_\times & 0\\
			0 & h_\times & - h_+ & 0\\
			0 & 0 & 0 & 0
		\end{array}
	\right) ,
\ee
where $h_+$ and $h_\times$ are two free functions of the retarded
time $t-z$. These are the \emph{polarization states} of the propagating
gravitational waves, the two radiative degrees of freedom of the
metric. So gravitational wave possess two linearly independent polarization states, just like electromagnetic waves. The reason why the polarizations are denoted $+$ and $\times$
will become clear in the next section, where we discuss the effect of
an incoming gravitational wave on matter.

In the TT gauge, gravitational waves are described by a $2 \times 2$ matrix in the plane orthogonal to the direction of propagation. Now, under a rotation of angle $\theta$ about that direction, the polarization states transform as
\be\label{e:helicity}
	h_+ \pm \ui \, h_\times \longrightarrow e^{\mp 2 \ui \theta} \left( h_+ \pm \ui \, h_\times \right) .
\ee
In the language of particle physics, the \emph{helicity} $\mathcal{H}$ of a particle is the projection of its spin along the direction of propagation. And, under a rotation of angle $\theta$ about that direction, the helicity states transform as $h \longrightarrow e^{\pm \ui \mathcal{H} \theta} h$. Therefore, \eref{e:helicity} shows that linearized gravity can be understood as the theory of a spin-2 particle with helicity states $h_+ \pm \ui \, h_\times$, the \emph{graviton}, just like the (massless) photon is a spin-1 particle responsible for mediating the electromagnetic interaction.

\section{Interaction of Gravitational Waves with Matter}\label{s:th_inter}

How can a gravitational wave be detected? In order to address this question, one must first understand how gravitational waves can interact with matter. Interestingly, this topic has historically been the source of heated debates.\cite{Ken}

\subsection{Description in the TT gauge}

We begin our analysis of this problem using the TT gauge introduced in the previous section. From the definition \eqref{e:def_h} and the expression \eqref{e:hijTT} for the metric perturbation, the spacetime interval in the TT gauge simply reads
\be
\label{e:ds2_TT_gauge}
  \ud s^2 = - \ud t^2 + \left( \delta_{ij} + h_{ij}^\text{TT} \right) \ud x^i \ud x^j \p
\ee
We first consider the motion of a free-falling test particle, which must obey the geodesic equation \eqref{e:geodesic2}. Notice that the proper time $\tau$ along that worldline coincides with the TT coordinate time $t$. By using the expression \eqref{e:def_Christoffel} for the Christoffel symbols with the metric components \eqref{e:ds2_TT_gauge}, one can show that the coordinate acceleration of a freely falling test mass vanishes:
\be
	\ddot{X}^i(t) = 0 \p
\ee
Therefore, if the particle was initially at rest, $\dot{X}^i(t_0) = 0$, it remains at rest with respect to the TT coordinates $(t,x^i)$. Beware, however, that this does not mean that gravitational waves have no effect on freely falling particles. Rather, the TT gauge is simply a coordinate system that is comoving with such particles.

The easiest way to understand the physical effects of gravitational waves on matter is to consider the \emph{relative} motion of two nearby test masses in free fall. Indeed, the distance $L$ between two such particles can be defined in an operational manner, by sending light rays back and forth and measuring the proper time elapsed between emission ($t=T_\text{em}$) and reception ($t=T_\text{rec}$): 
\be
	L \equiv \frac{1}{2} \left( T_\text{rec} - T_\text{em} \right) .
\ee
Recalling that light rays propagate along null geodesics, we can set $\ud s^2 = 0$ in Eq.~\eqref{e:ds2_TT_gauge} and obtain, to first order in the metric perturbation $h_{ij}^\text{TT}$,
\be\label{e:L(t)}
	L(t) = L_0 \left( 1 + \frac{1}{2} n^i n^j h_{ij}^\text{TT} \right) .
\ee
Here $L_0 \equiv \delta_{ij} \Delta x^i \Delta x^j$ is the (Euclidean) distance between the two masses in absence of gravitational wave and $n^i \equiv \Delta x^i / L_0$ a unit vector. Equation \eqref{e:L(t)} shows that the distance between the two particles varies in time under the effect of the propagating gravitational wave. It is essentially this change in the distance between test masses that existing gravitational wave detectors attempt to measure.

\subsection{Description using geodesic deviation}

In the TT gauge, there is close relationship between the metric perturbation $h_{\alpha\beta}^\text{TT}$ and the linearized Riemann tensor \eqref{e:linear_Riemann}, namely
\be\label{e:R_hTT}
	R_{itjt} = - \frac{1}{2} \ddot{h}_{ij}^\text{TT} \c
\ee
where the overdot stands for the partial derivative $\partial / \partial t$.
Now, recall that, in general relativity, all of the physical
effects of gravitation are encoded in the Riemann tensor. In
particular, as mentionned in Sec.~\ref{ss:Riemann}, the relative
acceleration of two neighboring geodesics is controlled by
$R_{abcd}$. Substituting for \eref{e:R_hTT} into the equation of
geodesic deviation \eqref{e:geo_dev} shows that, to first order in the
separation vector $s^a$, with components $s^\alpha = (0,\xi^i)$ and $u^\alpha = (1,\vec{0})$ in a local inertial frame (see below),
\be
	\frac{\ud^2 \xi^i}{\ud t^2} = \frac{1}{2} \ddot{h}_{ij}^\text{TT} \xi^j \p
\ee
Hence, the effect of a gravitational wave on matter can be understood as an additional Newtonian-like force, called a \emph{tidal force}, perturbing the relative acceleration between nearby freely falling particles.

\subsection{Description using Fermi coordinates}

As mentionned above, it is possible to give a quasi-Newtonian description of the motion of point masses under the action of gravitational radiation. To do so, one must introduce \emph{Fermi coordinates}, i.e., a local inertial frame defined in a neighborhood of an entire timelike geodesic, that deviates from the flat metric only quadratically in the distance from this worldline.

Thus, using Fermi coordinates $(\hat{x}^\alpha)$ defined in the vicinity of the worldline of a freely falling observer $\mathcal{O}$, the spacetime interval takes the form
\be\label{e:Fermi_coordinates}
	\ud s^2 = - {(\ud \hat{x}^0)}^2 + \delta_{ij} \, \ud \hat{x}^i \ud \hat{x}^j + \mathcal{O} \bigl( \Vert \hat{x}^i \Vert^2 / \mathcal{R}^2 \bigr) \c
\ee
where deviations from Minkowski's metric appear at quadratic order in the spatial distance $\Vert \hat{x}^i \Vert$ and $\mathcal{R}$ is the curvature radius such that $\vert R_{\alpha\beta\mu\nu} \vert \!\sim\! \mathcal{R}^{-2}$. The coordinate transformation from the TT gauge \eqref{e:ds2_TT_gauge} to the Fermi gauge \eqref{e:Fermi_coordinates} simply reads
\be\label{e:trans_TT_Fermi}
	\hat{x}^0 = t \c \quad \hat{x}^i = x^i + \frac{1}{2} h_{ij}^\text{TT}(t,\vec{0}) \, x^j \c
\ee
where $h_{ij}^\text{TT}(t,\vec{0})$ denotes the value of the field $h_{ij}^\text{TT}$ along the worldline of $\mathcal{O}$. (Here and in what follows we assume that the wavelength of the radiation is much larger than the typical size of the system of point masses.)

\begin{figure}
	\begin{center}
		\vspace{0.2cm}
		\includegraphics[width=10.7cm]{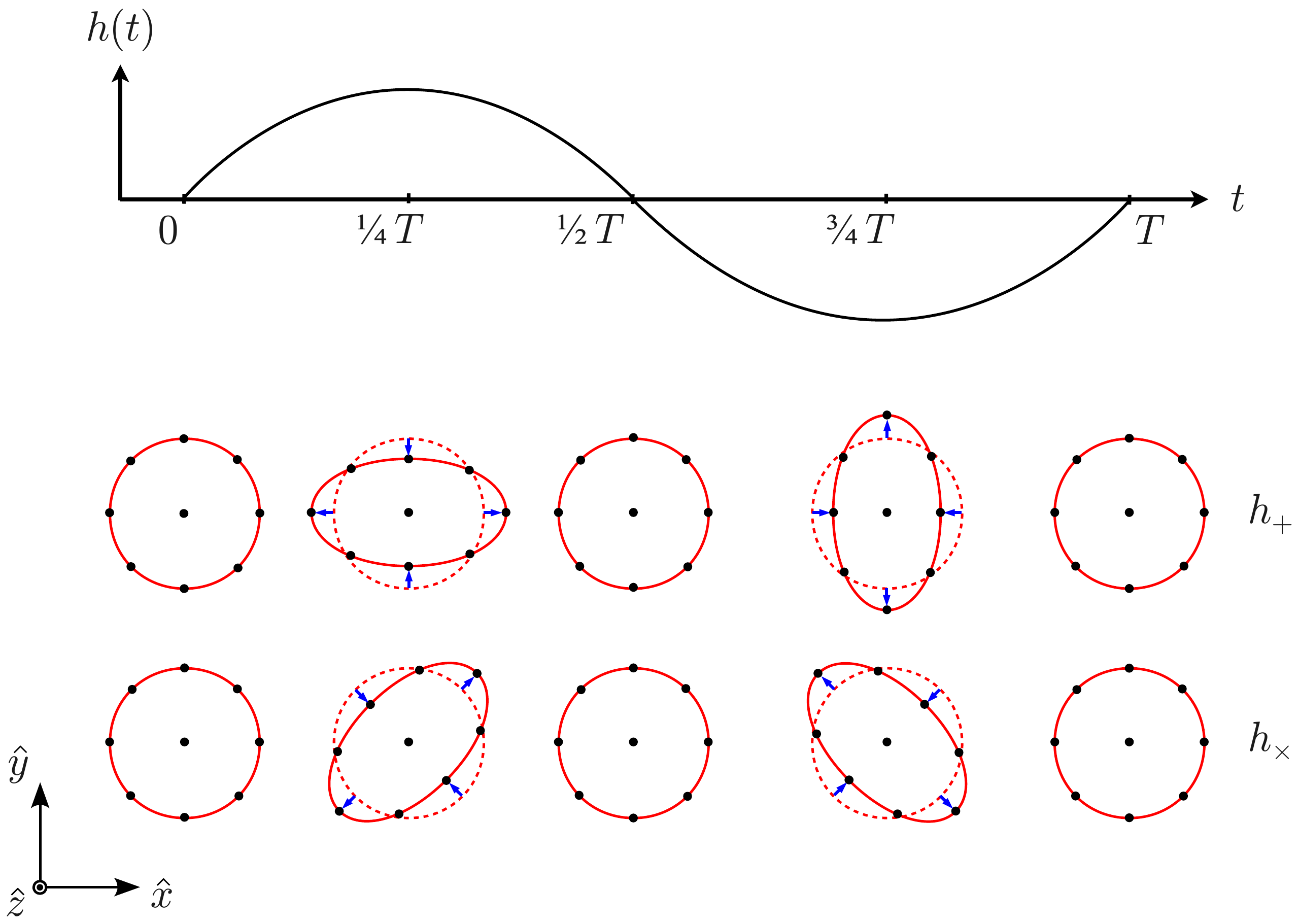}
		\caption{A monochromatic gravitational wave of pulsation $\omega = 2\pi/T$ propagates along the $\hat{z}$ direction. The lower panel shows the effects of the $+$ and $\times$ polarizations on a ring of freely falling particles, in a local inertial frame.}
		\label{f:polarizations}
	\end{center}
\end{figure}

Let us now consider a set of non-interacting ---i.e., freely falling--- point masses located in a neighborhood of $\mathcal{O}$. Since the spatial TT coordinates, say $x_0^i$, of one such particle do not change as the gravitational wave passes, Eq.~\eqref{e:trans_TT_Fermi} implies that its trajectory in the Fermi coordinates associated to the observer $\mathcal{O}$ is given by
\be\label{e:motion_Fermi}
	\hat{x}^i(t) = x_0^i + \frac{1}{2} h_{ij}^\text{TT}(t,\vec{0}) \, x_0^j \p
\ee
This formula can be applied to the particular case of a monochromatic wave of pulsation $\omega = 2\pi / T$ (as measured by the observer) that propagates along the $\hat{z}$ direction. Using \eref{e:hijTT} with $h_{+,\times}(t) = H_{+,\times} \, e^{\ui \omega t}$, this gives 
\begin{subequations}\label{e:move_dust}
	\begin{align}
		\hat{x}(t) &= x_0 + \frac{1}{2} \left( H_+ x_0 + H_\times y_0 \right) e^{\ui \omega t} \c \\
		\hat{y}(t) &= y_0 + \frac{1}{2} \left( H_\times x_0 - H_+ y_0 \right) e^{\ui \omega t} \c \\
		\hat{z}(t) &= z_0 \p
	\end{align}
\end{subequations}
Thus, as a gravitational wave propagates through an initially circular ring of particles, it induces alternative contractions and elongations along the $\hat{x}$ and $\hat{y}$ directions for the $+$ polarization, and along the $\hat{y} = \hat{x}$ and $\hat{y} = -\hat{x}$ directions for the $\times$ polarization (see Fig.~\ref{f:polarizations}). A generic gravitational wave can thus be understood as a superposition of two \emph{oscillating tidal fields} that propagate at the vacuum speed of light. 

Equation \eqref{e:move_dust} shows that under the effect of a passing gravitational wave of typical amplitude $h \sim H_{+,\times}$, the initial size $L_0$ of the ring of particles varies by an amount
\be
	\delta L \sim \frac{1}{2} \, h \, L_0 \c
\ee
in complete agreement with the result \eqref{e:L(t)}. As will be shown in section \ref{s:gen}, the typical amplitude of gravitational waves from astrophysical sources is $h \lesssim 10^{-21}$. Hence, even for a kilometer-scale detector, the change in length induced by a traveling gravitational wave is at most of order $10^{-18}\,\text{m}$. Thus, as will be discussed in chapters 3 and 4, it is a major technological challenge to detect a passing gravitational wave of cosmic origin.

\section{Generation of Gravitational Waves}\label{s:gen}

In the previous section, we have seen how propagating gravitational waves can affect the motion of matter. In this section we will show how, conversely, the motion of matter generates gravitational radiation.
  
\subsection{Einstein's quadrupole formula}\label{ss:th_QF}

\begin{figure}
	\begin{center}
		\vspace{0.8cm}
		\includegraphics[width=12cm]{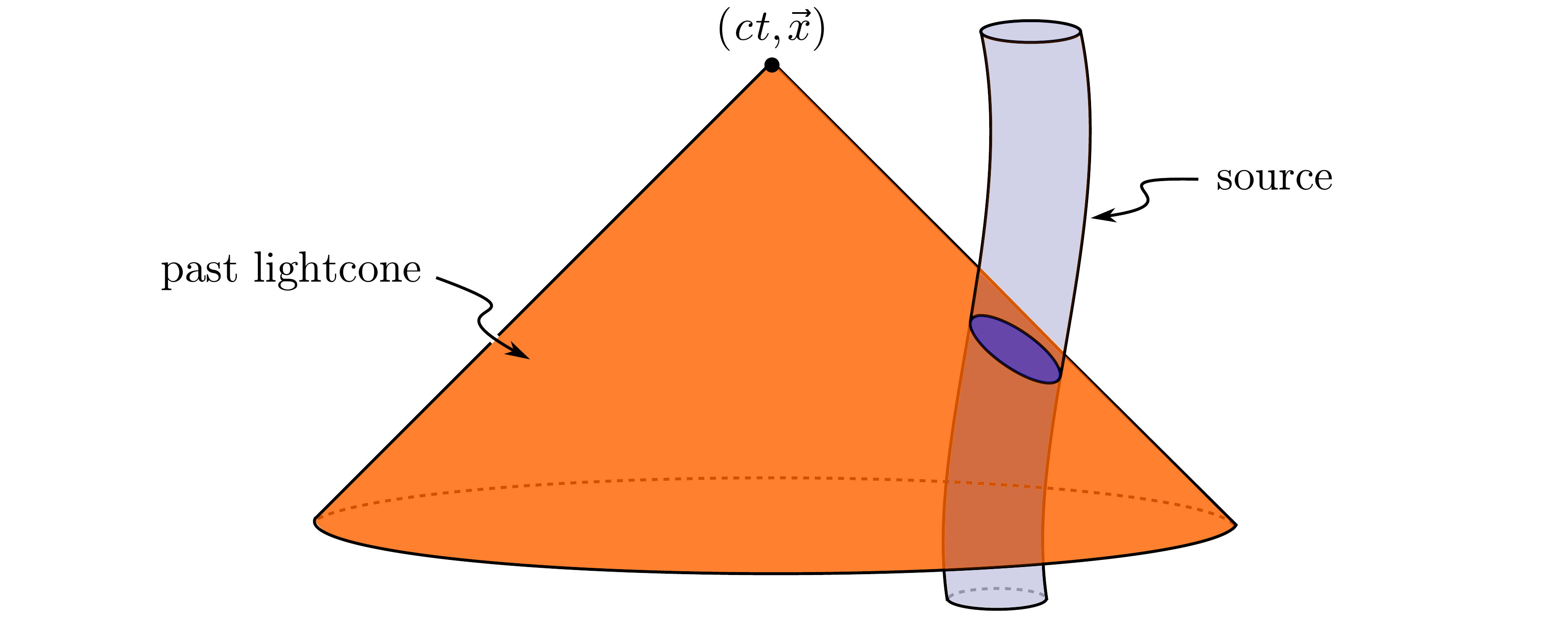}
		\caption{Gravitational waves propagate at the finite velocity $c$. Hence the field at a point $(ct,\vec{x})$ is only influenced by the matter source at the retarded times $t' = t - \Vert \vec{x} - \vec{y} \Vert / c$.}
		\label{f:source}
	\end{center}
\end{figure}

We shall describe the generation of gravitational waves by \emph{isolated} systems, and we consider again the linearized version \eqref{e:box_h} of Einstein's equation, in the harmonic gauge. Therefore, the following discussion will be restricted to the simplest case of \emph{weakly relativistic} sources, for which the linearized approximation is valid. Now, for each component $\bar{h}_{\alpha\beta}$ of the perturbation $\bar{h}_{ab}$, the linear wave equation \eqref{e:box_h} can be solved using the standard formula for retarded potentials, namely (restoring powers of $1/c$)
\be
	\bar{h}_{\alpha\beta}(t,\vec{x}) = \frac{4G}{c^4} \int_{\mathbb{R}^3} \frac{T_{\alpha\beta}(t',\vec{y})}{\Vert \vec{x} - \vec{y} \Vert} \, \ud^3 y \p
\ee
This is an integral over the past lightcone of the field point $(ct,\vec{x})$. Unlike in Newtonian gravity, the gravitational field at a point $(ct,\vec{x})$ is only influenced by the matter source at the \emph{retarded} times $t' \equiv t - \Vert \vec{x} - \vec{y} \Vert / c$, the lag resulting from the time needed for a signal propagating at the speed of light $c$ to get from a point $\vec{y}$ inside the source to the point $\vec{x}$ (see \fref{f:source}).

Assuming in addition that the source is \emph{slowly varying}, in the sense that its energy-momentum tensor does not vary much over a light-crossing time, standard manipulations yield for the field in the radiation zone
\be\label{e:h_intT}
	\bar{h}_{\alpha\beta}(t,\vec{x}) = \frac{4G}{c^4 r} \int_{\mathbb{R}^3} T_{\alpha\beta}(t - r/c, \vec{y}) \, \ud^3 y \c
\ee
where $r = \Vert \vec{x} \Vert$ is the distance from the source and all terms of $\calO(r^{-2})$ have been neglected. As already mentioned, the radiative degrees of freedom are contained in the spatial part of the metric. Then, by combining \eqref{e:h_intT} with the conservation of the energy-momentum tensor for a ball of perfect fluid in linearized gravity, i.e., $\partial^\alpha T_{\alpha\beta} = 0$ where $T_{\alpha\beta}$ is given by \eref{e:def_perfect_fluid} above, it can be shown that the spatial components of $\bar{h}_{\alpha\beta}$ are given by
\be\label{e:quadrupole0}
	\bar{h}_{ij}(t,\vec{x}) = \frac{2G}{c^4 r} \, \ddot{I}_{ij}(t - r/c) \p
\ee
Here the overdot stands for a derivative with respect to time, and the tensor $I_{ij}$ is the moment of inertia of the source $\mathcal{S}$, given by the following volume integral over the Newtonian mass density $\rho$:
\be\label{e:def_Iij}
	I_{ij}(t) = \int_\mathcal{S} \rho(t,\vec{x}) \, x_i x_j \, \ud^3 x \p
\ee

Now, to obtain the metric perturbation in the TT gauge, it is enough to consider the transverse-traceless part of Eq.~\eqref{e:quadrupole0}. This is achieved by means of an appropriate projection. First, however, we introduce the traceless part of the tensor \eqref{e:def_Iij}, the source's \emph{mass quadrupole moment}
\be\label{e:mass_quadrupole}
	Q_{ij}(t) = \int_\mathcal{S} \rho(t,\vec{x}) \, \Bigl( x_i x_j - \frac{1}{3} \Vert \vec{x} \Vert^2 \delta_{ij} \Bigr) \, \ud^3 x \c
\ee
a quantity that is directly related to the multipolar expansion of the Newtonian gravitational potential of the source: $\Phi = - \frac{GM}{r} + \frac{3G}{2r^3} \, Q_{ij} n^i n^j + \cdots$, with $M$ the mass of the source and $\vec{n} = \vec{x} / r$ the unit direction from the origin. At last, Einstein's famous \emph{quadrupole formula} simply reads
\be\label{e:quadrupole}
	h_{ij}^\text{TT}(t,\vec{x}) = \frac{2G}{c^4 r} \, \Lambda_{ij,kl}(\vec{n}) \, \ddot{Q}_{kl}(t - r/c) \c
\ee
where $\Lambda_{ij,kl} \equiv P_{ik} P_{jl} - \frac{1}{2} P_{ij} P_{kl}$ is defined in terms of the projection operator $P_{ij} \equiv \delta_{ij} - n_i n_j$ orthogonal to the direction of propagation.

This key result shows that, to leading order in a multipolar expansion, gravitational waves are generated by any time-varying quadrupole moment. The laws of conservation of mass and linear momentum forbid the emission of monopolar or dipolar gravitational radiation. In electromagnetism, while the electric charge (the monopole) is conserved, the electric dipole moment is not, so electromagnetic radiation is predominantly dipolar.

Although the quadrupole formula \eqref{e:quadrupole} is valid for sources whose dynamics is dominated by gravitational forces, the derivation we sketched above is not. Indeed, in linearized gravity the exact conservation law \eqref{e:div_T} reduces to $\partial_a T^{ab} = 0$, such that all bodies move along geodesics of Minkowski spacetime, thereby excluding gravitationally bound orbits. The extension of this derivation to the case of sources with nonnegligible self-gravity is important for computing the gravitational-wave emission to be expected from binary
systems of compact objects, whose orbits become highly relativistic just before coalescence, and which cannot be described by the linearized theory; see for instance Refs.~\refcite{Bl.14,BuSa.15} and chapter 2 in this book.

The quadrupole formula \eqref{e:quadrupole} can be used to get an order-of-magnitude estimate of the amplitude $h$ of gravitational waves generated by a source of mass $M$, typical size $R$ and quadrupole moment $Q \sim s M R^2$, where $0 \leq s \lesssim 1$ is an asymmetry fudge factor, such that $s=0$ for a spherically symmetric source. If $\omega$ denotes the inverse of the timescale of evolution of the source ---the angular velocity for a quasi-periodic source---then $\ddot{Q} \sim s \, \omega^2 M R^2$ and \eref{e:quadrupole} yields
\be\label{e:h}
	h \sim \frac{2G}{c^4r} \, \omega^2 s M R^2 \sim \frac{R}{r} \, \frac{R_\text{S}}{R} \left( \frac{v}{c} \right)^2 s \c
\ee
where we introduced the Schwarzschild radius $R_\text{S} \equiv 2GM / c^2$ and the typical internal velocity $v \sim R \, \omega$ of the source. Equation \eqref{e:h} shows that in the most favorable case of a nonspherical ($s \lesssim 1$) and compact source ($R \gtrsim R_\text{S}$) moving at relativistic speeds ($v \lesssim c$), we get $h \lesssim R / r \lesssim GM/(c^2r)$. For a $3 M_\odot$ source located at $200~\text{Mpc}$, for instance the coalescence of two neutron stars in a distant galaxy, this gives the estimate $h \lesssim 10^{-22}$.

\subsection{Gravitational luminosity}

If the typical wavelength $\lambda$ of gravitational waves is much smaller than the characteristic radius of curvature $\mathcal{R}$ of the background spacetime, then the separation of scales can be used to introduce an \emph{effective} energy-momentum tensor associated with the gravitational radiation. Its expression is given by the average $\langle \cdot \rangle$ over several wavelengths of the second-order contributions in the expansion of the Einstein tensor in powers of the metric perturbation (\sref{ss:linear}), namely\cite{Is1.68,Is2.68}
\be\label{e:def_Isaacson}
  T_{ab} = \frac{c^4}{32\pi G} \, \Big\langle \partial_a \bar{h}_{cd} \, \partial_b \bar{h}^{cd} - \frac{1}{2} \partial_a \bar{h} \, \partial_b \bar{h} - 2 \partial_{(a} \bar{h}_{b)c} \, \partial_d \bar{h}^{cd} \Big\rangle \p
\ee
This \emph{Isaacson energy-momentum tensor} effectively localizes the energy and momentum content in short-wavelength gravitational radiation over regions whose size is comparable to $\lambda \ll \mathcal{R}$. It can be checked that the right-hand side of \eref{e:def_Isaacson} is invariant under gauge transformations of the form \eqref{e:change_h}. In the TT gauge, in which $\bar{h} = 0$ and $\partial_d \bar{h}^{cd} = 0$, its coordinate components simply reduce to
\be\label{e:def_Isaacson_TT}
	T_{\alpha\beta} = \frac{c^4}{32\pi G} \, \big\langle \partial_\alpha h^\text{TT}_{\mu\nu} \partial_\beta h_\text{TT}^{\mu\nu} \big\rangle \p
\ee

For gravitational radiation that propagates along the $z$-axis, the flux of energy $F$ carried by the waves is given by the component $T_{tz}$ of the energy- momentum tensor \eqref{e:def_Isaacson_TT}; recall Sec.~\ref{ss:Tmunu}. Then, using Eq.~\eqref{e:hijTT} one finds
\be\label{e:GW_flux}
	F = \frac{c^3}{16\pi G} \, \big\langle \dot{h}^2_+ + \dot{h}^2_\times \big\rangle \p 
\ee
For gravitational waves with a typical frequency $f$ and amplitude $h$, \eref{e:GW_flux} implies $F \sim c^3 f^2 h^2 / (32\pi G)$. Now, for realistic astrophysical sources, such as neutron star binaries,\footnote{The typical gravitational-wave frequency of a source of mass $M$, linear size $R \gtrsim GM/c^2$ and mean density $\bar{\rho} \sim M / R^3 \lesssim c^6 / (G^3 M^2)$ is $f \sim \sqrt{G \bar{\rho}} \lesssim c^3 / (GM)$; see chapter 2.} $f \sim 1~\text{kHz}$ and $h \sim 10^{-22}$ yields $F \sim 3~\text{mW} \cdot \text{m}^{-2}$. Hence, a gravitational wave with a tiny amplitude can carry a large amount of energy. By analogy with the theory of elasticity, spacetime can be though of as an extremely ``rigid medium.''

Integrating the energy flux \eqref{e:GW_flux} over a 2-sphere whose radius $r$ is taken to infinity, one obtains the \emph{gravitational luminosity} of a given source as
\be\label{e:GW_luminosity0}
	L = \lim_{r \to \infty} \frac{c^3 r^2}{16\pi G} \int \big\langle \dot{h}_+^2 + \dot{h}_\times^2 \big\rangle \, \ud \Omega \c
\ee
where $\ud \Omega$ stands for the surface element of the unit 2-sphere. Substituting for the first Einstein quadrupole formula, \eref{e:quadrupole}, into \eqref{e:GW_luminosity0} yields the total power radiated as a function of the source quadrupole moment:
\be\label{e:GW_luminosity}
	L = \frac{G}{5 c^5} \, \big\langle \dddot{Q}_{ij} \dddot{Q}_{ij} \big\rangle \p
\ee
This is Einstein's second \emph{quadrupole formula}. Interestingly, this expression can be compared to a similar result established in electromagnetism, where it can be shown that the power radiated by a slowly-varying distribution of accelerated charges with dipole moment $D_i$ reads [with $\mu_0 = 1/(\varepsilon_0 c^2)$]
\be
	L_\text{e.m.} = \frac{2}{3} \frac{\mu_0}{4 \pi c} \, \big\langle \ddot{D}_i \ddot{D}_i \big\rangle \p
\ee

Equation \eqref{e:GW_luminosity} can be used to get an order-of-magnitude estimate of the gravitational luminosity of a source of mass $M$ and typical size $R$, for which $Q \sim s M R^2$. Again, if $\omega$ denotes the inverse of the timescale of evolution of the source, then $\dddot{Q} \sim s \, \omega^3 M R^2$ and \eref{e:GW_luminosity} yields
\be\label{e:oom_luminosity}
	L \sim \frac{G}{c^5} \, s^2 \omega^6 M^2 R^4 \sim \frac{c^5}{4G} \left( \frac{R_\text{S}}{R} \right)^2 \left( \frac{v}{c} \right)^6 s^2 \p
\ee
This formula clearly shows that a Hertz-type experiment is hopeless; no laboratory experiment will ever produce a significant amount of gravitational radiation that can be detected on Earth. However, nonspherical $(s \lesssim 1$) and compact objects ($R \gtrsim R_\text{S}$) moving at relativistic speeds ($v \lesssim c$) are powerful gravitational-wave sources, with $L \lesssim c^5 / 4G \simeq 10^{52}~\text{W}$. By comparison, the luminosity of the Sun is a mere $3.8 \times 10^{26}~\text{W}$, that of a typical galaxy is of the order of $10^{37}~\text{W}$, while all the galaxies in the visible Universe emit, in visible light, of the order of $10^{49}~\text{W}$. Binary black hole mergers can thus, at the peak of their wave emission, compete in luminosity with the steady luminosity of the entire Universe! For instance, the binary black hole merger event GW150914 radiated about $3M_\odot c^2$ of energy within $250~\text{ms}$, reaching a peak emission rate of $3.6 \times 10^{49}~\text{W}$, which is equivalent to $200 M_\odot c^2/\text{s}$.\cite{Ab.al2.16}

\bibliographystyle{ws-rv-van}
\bibliography{/home/letiec/ownCloud/Publications/ListeRef}

\begin{thebibliography}{42}
\providecommand{\natexlab}[1]{#1}
\providecommand{\url}[1]{\texttt{#1}}
\expandafter\ifx\csname urlstyle\endcsname\relax
  \providecommand{\doi}[1]{doi: #1}\else
  \providecommand{\doi}{doi: \begingroup \urlstyle{rm}\Url}\fi

\bibitem{Ei.16}
A.~Einstein, N{\"a}herungsweise integration der feldgleichungen der
  gravitation, \emph{Sitzber. Preuss. Akad. Wiss.} p. 688  (1916).

\bibitem{Ei.18}
A.~Einstein, Gravitationswellen, \emph{Sitzber. Preuss. Akad. Wiss.} p. 154
  (1918).

\bibitem{HuTa.75}
R.~A. Hulse and J.~H. Taylor, Discovery of a pulsar in a binary system,
  \emph{Astrophys. J.} {\bf 195}, \penalty0 L51  (1975).

\bibitem{WeHu.16}
J.~M. Weisberg and Y.~Huang, Relativistic measurements from timing the binary
  pulsar {PSR B1913+16}, \emph{Astrophys. J.}  (2016).

\bibitem{Lo.08}
D.~R. Lorimer, Binary and millisecond pulsars, \emph{Living Rev. Relativity}.
  {\bf 11}, \penalty0 8  (2008).

\bibitem{Aa.al.15}
J.~Aasi et~al., Advanced {LIGO}, \emph{Class. Quant. Grav.} {\bf 32}, \penalty0
  074001  (2015).

\bibitem{Ac.al.15}
F.~Acernese et~al., Advanced {V}irgo: a second-generation interferometric
  gravitational wave detector, \emph{Class. Quant. Grav.} {\bf 32}, \penalty0
  024001  (2015).

\bibitem{Ab.al2.16}
{B.~P.~Abbott et al. (LIGO Scientific Collaboration and Virgo Collaboration)},
  Observation of gravitational waves from a binary black hole merger,
  \emph{Phys. Rev. Lett.} {\bf 116}, \penalty0 061102  (2016).

\bibitem{Ab.al3.16}
{B.~P.~Abbott et al. (LIGO Scientific Collaboration and Virgo Collaboration)},
  {GW151226}: Observation of gravitational waves from a 22-solar-mass binary
  black hole coalescence, \emph{Phys. Rev. Lett.} {\bf 116}, \penalty0 241103
  (2016).

\bibitem{Ab.al.16}
{B.~P.~Abbott et al. (LIGO Scientific Collaboration and Virgo Collaboration)},
  Prospects for observing and localizing gravitational-wave transients with
  {A}dvanced {LIGO} and {A}dvanced {V}irgo, \emph{Living Rev. Relativity}. {\bf
  19}, \penalty0 1  (2016).

\bibitem{SaSc.09}
B.~S. Sathyaprakash and B.~F. Schutz, Physics, astrophysics and cosmology with
  gravitational waves, \emph{Living Rev. Relativity}. {\bf 12}, \penalty0 2
  (2009).

\bibitem{Ce.03}
J.~M. Centrella, Resource letter: Gravitational waves, \emph{Am. J. Phys.} {\bf
  71}, \penalty0 520  (2003).

\bibitem{Th.87}
K.~S. Thorne.
\newblock Gravitational radiation.
\newblock In eds. S.~W. Hawking and W.~Israel, \emph{Three hundred years of
  gravitation}, p. 330, Cambridge University Press, Cambridge  (1987).

\bibitem{ScRi.01}
B.~F. Schutz and F.~Ricci.
\newblock Gravitational waves, sources and detectors.
\newblock In eds. I.~Ciufolini, V.~Gorini, U.~Moschella, and P.~Fr{\'e},
  \emph{Gravitational waves}, p.~11, Institute of Physics Publishing  (2001).

\bibitem{FlHu.05}
{\'E}.~{\'E}. Flanagan and S.~A. Hughes, The basics of gravitational wave
  theory, \emph{New J. Phys.} {\bf 7}, \penalty0 204  (2005).

\bibitem{Bu.07}
A.~Buonanno.
\newblock Gravitational waves.
\newblock In eds. F.~Bernardeau, C.~Grojean, and J.~Dalibard, \emph{Particle
  physics and cosmology: The fabric of spacetime}, vol.~86, \emph{Les Houches},
  p.~3, Elsevier  (2007).

\bibitem{Bl.14}
L.~Blanchet, Gravitational radiation from post-{N}ewtonian sources and
  inspiralling compact binaries, \emph{Living Rev. Relativity}. {\bf 17},
  \penalty0 2  (2014).

\bibitem{BuSa.15}
A.~Buonanno and B.~S. Sathyaprakash.
\newblock Sources of gravitational waves: Theory and observations.
\newblock In eds. A.~Ashtekar, B.~K. Berger, J.~Isenberg, and M.~{MacCallum},
  \emph{General relativity and gravitation: A centennial perspective}, p. 287,
  Cambridge University Press, Cambridge  (2015).

\bibitem{Mag}
M.~Maggiore, \emph{Gravitational waves: Theory and experiments}. Oxford
  University Press, Oxford  (2007).

\bibitem{CrAn}
G.~D.~E. Creighton and W.~G. Anderson, \emph{Gravitational-wave physics and
  astronomy: An introduction to theory, experiment and data analysis}.
  Wiley-VCH, Weinheim  (2011).

\bibitem{Wei}
S.~Weinberg, \emph{Gravitation and cosmology}. John Wiley, New York  (1972).

\bibitem{MTW}
C.~W. Misner, K.~S. Thorne, and J.~A. Wheeler, \emph{Gravitation}. Freeman, New
  York  (1973).

\bibitem{Wal}
R.~M. Wald, \emph{General relativity}. University of Chicago Press, Chicago
  (1984).

\bibitem{Har}
J.~B. Hartle, \emph{Gravity: An introduction to {E}instein's general
  relativity}. Addison Wesley, San Fransisco  (2003).

\bibitem{Car}
S.~M. Carroll, \emph{Spacetime and geometry: An introduction to general
  relativity}. Addison Wesley, San Fransisco  (2004).

\bibitem{Schu}
B.~Schutz, \emph{A first course in general relativity}. Cambridge University
  Press, Cambridge  (2009).

\bibitem{Str}
N.~Straumann, \emph{General relativity}, second edn. Springer, New York
  (2013).

\bibitem{Gou2}
E.~Gourgoulhon, \emph{Special relativity in general frames}. Graduate Texts in
  Physics, Springer, New York  (2013).

\bibitem{Ra.04}
F.~Ravndal.
\newblock Scalar gravitation and extra dimensions.
\newblock In eds. C.~Cronstr{\"o}m and C.~Montonen, \emph{Proceedings of the
  {G}unnar {N}ordstr{\"o}m symposium on theoretical physics}, p. 151, Finnish
  Society of Sciences and Letters, Helsinki  (2004).

\bibitem{An.al.13}
S.~Ando et~al., Multimessenger astronomy with gravitational waves and
  high-energy neutrinos, \emph{Rev. Mod. Phys.} {\bf 85}, \penalty0 1401
  (2013).

\bibitem{Am.al3.13}
L.~Amati et~al.
\newblock Light from the cosmic frontier: Gamma-ray bursts  (2013).

\bibitem{Ot.09}
C.~D. Ott, The gravitational-wave signature of core-collapse supernovae,
  \emph{Class. Quant. Grav.} {\bf 26}, \penalty0 063001  (2009).

\bibitem{Wi.14}
C.~M. Will, The confrontation between general relativity and experiment,
  \emph{Living Rev. Relativity}. {\bf 17}, \penalty0 4  (2014).

\bibitem{Eo.al.22}
R.~V. E{\"o}tv{\"o}s, D.~Pek{\'a}r, and E.~Fekete, Beitr{\"a}ge zum gesetze der
  proportionalit{\"a}t von tr{\"a}gheit und gravit{\"a}t, \emph{Ann. Phys.}
  {\bf 373}, \penalty0 11  (1922).

\bibitem{Sc.al2.08}
S.~Schlamminger, K.-Y. Choi, T.~A. Wagner, J.~H. Gundlach, and E.~G.
  Adelberger, Test of the equivalence principle using a rotating torsion
  balance, \emph{Phys. Rev. Lett.} {\bf 100}, \penalty0 041101  (2008).

\bibitem{To.al.12}
P.~Touboul, G.~M{\'e}tris, V.~Lebat, and A.~Robert, The {MICROSCOPE}
  experiment, ready for the in-orbit test of the equivalence principle,
  \emph{Class. Quant. Grav.} {\bf 29}, \penalty0 184010  (2012).

\bibitem{Mu.al.12}
J.~M{\"u}ller, F.~Hofmann, and L.~Biskupek, Testing various facets of the
  equivalence principle using lunar laser ranging, \emph{Class. Quant. Grav.}
  {\bf 29}, \penalty0 184006  (2012).

\bibitem{Go.al.11}
M.~E. Gonzalez et~al., High-precision timing of five millisecond pulsars: Space
  velocities, binary evolution, and equivalence principles, \emph{Astrophys.
  J}. {\bf 743}, \penalty0 102  (2011).

\bibitem{Ra.al.14}
S.~M. Ransom et~al., A millisecond pulsar in a stellar triple system,
  \emph{Nature}. {\bf 505}, \penalty0 520  (2014).

\bibitem{Ken}
D.~J. Kennefick, \emph{Traveling at the speed of thought: Einstein and the
  quest for gravitational waves}. Princeton University Press, Princeton
  (2007).

\bibitem{Is1.68}
R.~A. Isaacson, Gravitational radiation in the limit of high frequency.
  \textsc{i}. {T}he linear approximation and geometrical optics, \emph{Phys.
  Rev.} {\bf 166}, \penalty0 1263  (1968).

\bibitem{Is2.68}
R.~A. Isaacson, Gravitational radiation in the limit of high frequency.
  \textsc{ii}. {N}onlinear terms and the effective stress tensor, \emph{Phys.
  Rev.} {\bf 166}, \penalty0 1272  (1968).

\end{thebibliography}

\end{document}